\documentclass[journal,10pt]{IEEEtran}

\usepackage{amsmath,amssymb}
\usepackage{subfigure}
\usepackage{graphicx,graphics,color,psfrag}
\usepackage{cite,balance}
\usepackage{algorithm}
\usepackage{accents}
\usepackage{amsthm}
\usepackage{bm} 
\usepackage{url}
\usepackage{algorithmic}
\usepackage[english]{babel}
\usepackage{multirow}
\usepackage{enumerate}
\usepackage{cases}
\usepackage{stfloats}
\usepackage{dsfont}
\usepackage{color,soul}
\usepackage{amsfonts}
\usepackage{tcolorbox}

\usepackage{booktabs} % For better table formatting
\usepackage{array} 

\usepackage{algorithm}

\usepackage{acronym} % newly added.

\usepackage{makecell}

\usepackage{cite,graphicx,amsmath,amssymb}
\usepackage{fancyhdr}
\usepackage{hhline}
\usepackage{graphicx,graphics}
\usepackage{array,color}
\usepackage{amsmath}
\usepackage{amsthm}
\usepackage[flushleft]{threeparttable}
\IEEEoverridecommandlockouts

\newtheorem{remark}{Remark}

\include{header}
\newcommand{\mv}[1]{\mbox{\boldmath{$ #1 $}}}

\newcommand{\mr}[1]{\mathrm{#1}}

\makeatletter
\def\endthebibliography{%
	\def\@noitemerr{\@latex@warning{Empty `thebibliography' environment}}%
	\endlist
}
\makeatother

\begin{document}
	\title{Hierarchically Tunable 6DMA for Wireless Communication and Sensing: Modeling and Performance Optimization} %\IEEEspecialpapernotice{(Invited Paper)}
	\author{Haocheng Hua, \textit{Member, IEEE}, Yuyan Zhou, Weidong Mei, \textit{Member, IEEE}, Jie Xu, \textit{Fellow, IEEE}, Rui Zhang, \textit{Fellow, IEEE} \\
%		\thanks{Part of this paper has been submitted to IEEE International Conference on Communications (ICC), Montreal, Canada, June 8-12, 2025 \cite{hua2021transmit}.}
		\thanks{
			H. Hua and Y. Zhou with the School of Science and Engineering, The Chinese University of Hong Kong (Shenzhen), Guandong 518172, China (e-mail: huahaocheng@cuhk.edu.cn, yuyanzhou1@link.cuhk.edu.cn).}
		\thanks{W. Mei are with the National Key Laboratory of Wireless Communications, University of Electronic Science and Technology of China, Chengdu 611731, China (e-mail: wmei@uestc.edu.cn).} % checked at 1/7/2025 to be member.
		\thanks{J. Xu is with the School of Science and Engineering (SSE), the Shenzhen Future Network of Intelligence Institute (FNii-Shenzhen), and the Guangdong Provincial Key Laboratory of Future Networks of Intelligence, The Chinese University of Hong Kong (Shenzhen), Guandong 518172, China (email: xujie@cuhk.edu.cn).}
		\thanks{R. Zhang is with the School of Science and Engineering, Shenzhen Research Institute of Big Data, The Chinese University of Hong Kong (Shenzhen), Guangdong 518172, China. He is also with the Department of Electrical and Computer Engineering, National University of Singapore, Singapore 117583 (e-mail: elezhang@nus.edu.sg). R. Zhang is the corresponding author.}
	}
	
	\maketitle
	
	\begin{abstract}
		This paper proposes a new hierarchically tunable six-dimensional movable antenna (HT-6DMA) architecture for base station (BS) in future wireless networks, aiming to improve the performance of both wireless communication and sensing. 
		The HT-6DMA BS consists of multiple antenna arrays that can flexibly move on a spherical surface, with their three-dimensional (3D) positions and 3D rotations/orientations efficiently characterized
		in the global spherical coordinate system (SCS) and their individual local SCSs, respectively. 
		As a result, the 6DMA system is hierarchically tunable in the sense that each array's global position and local rotation can be separately adjusted in a sequential manner with the other being fixed, thus greatly reducing their design complexity and improving the achievable performance.
		In particular, we consider an HT-6DMA BS serving multiple single-antenna users in the uplink communication or sensing potential unmanned aerial vehicles (UAVs)/drones in a given airway area.
		Specifically, for the communication scenario, we aim to maximize the average sum rate of communication users in the long term by optimizing the positions and rotations of all 6DMA arrays at the BS.
		While for the airway sensing scenario, we maximize the minimum received sensing signal power along the airway by optimizing
		the 6DMA arrays' positions and rotations along with the BS's transmit covariance matrix.
		Despite that the formulated problems are both non-convex and challenging to solve, we propose efficient solutions to them by exploiting the hierarchical tunability of positions/rotations of 6DMA arrays in our proposed model.
		Numerical results show that the proposed HT-6DMA design significantly outperforms not only the traditional BS with fixed-position antennas (FPAs), but also the existing 6DMA
		scheme based on alternating array position/rotation optimization.
		%		 in the communication scenario in terms of solution quality and convergence speed. % Ok, Let's add 6DMA also in sensing.
		Furthermore, it is unveiled that the performance gains of HT-6DMA mostly come from the arrays' global position adjustments on the spherical surface, rather than their local rotation adjustments, which provides a useful guide for implementing 6DMA systems under practical performance-complexity trade-off consideration.
	\end{abstract}
	
	\section{Introduction} % make it single page length
	\label{sec:intro}
	
	Six-generation (6G) wireless networks  
	have attracted tremendous research interests from both industry and academia \cite{wang2023road,IMT_2030_6G_vision_new,du2024overview,lu2024tutorial}. Besides providing enhanced communication capabilities beyond fifth-generation (5G) networks, 6G aims to support brand new usage scenarios including artificial intelligence (AI) and communication, integrated sensing and communication (ISAC) \cite{liu2022integrated}, and ubiquitous connectivity \cite{IMT_2030_6G_vision_new}. In particular,  
	the multiple-input-multiple-output (MIMO) technology is a key enabler for 6G \cite{lu2024tutorial,liu2022integrated}, which provides not only spatial multiplexing and diversity gains to significantly enhance the efficiency and reliability of wireless communication \cite{heath2018foundations}, but also waveform and spatial diversity gains to substantially improve the accuracy and resolution of wireless sensing \cite{li2008mimo}. 
	Moreover, MIMO techniques have been widely used in ISAC to enable simultaneous multi-target sensing and multi-user communications \cite{hua2023optimal,hua2022mimojournal,hua2024near}. 
	As a result, increasing the number of antennas deployed in wireless networks in a centralized or decentralized manner has been recognized as an efficient solution to further improve the performance of MIMO systems \cite{lu2024tutorial}, and various advanced MIMO systems such as hybrid analog/digital beamforming \cite{sohrabi2016hybrid}, holographic MIMO \cite{huang2020holographic}, as well as dynamic metasurface antennas (DMAs) \cite{shlezinger2021dynamic} have been proposed to reduce the hardware cost and processing complexity of large-scale MIMO systems with enhanced performance.
	In addition, intelligent reflecting surface (IRS) or reconfigurable intelligent surface (RIS)-assisted MIMO has been proposed to reconfigure wireless channels by smart signal reflection/refraction for enhancing sensing \cite{shao2022target} and communication \cite{wu2024intelligent} performances  with low cost and energy consumption. 
	
	However, the aforementioned MIMO techniques rely on fixed-position antennas (FPAs) with non-adjustable antenna positions after deployment, which cannot 
	adapt to time/space-varying wireless channels and performance requirements. To overcome this drawback, movable antenna (MA) has been introduced to wireless networks\footnote{MA is also known as fluid antenna in terms of antenna position adjustment \cite{zhu2024historical}.}, which enables the adjustment of antenna positions according to instantaneous/statistical channels as well as specific  tasks with different design objectives \cite{zhu2023modeling,zhu2024performance,mei2024movable,ma2023mimo,zhu2023movablenull,wei2024joint,zhu2023movable,ma2024movable,chen2024exploiting, xiao2024channel,li2024minimizing,wang2024throughput}. 
	Prior works	have demonstrated the advantages of MAs over FPAs for wireless communication, showing their superiority in enhancing channel gains \cite{zhu2023modeling,zhu2024performance,mei2024movable} and MIMO channel capacity \cite{ma2023mimo},
	flexible beamforming \cite{zhu2023movablenull}, interference mitigation \cite{wei2024joint}, and transmit power minimization \cite{zhu2023movable}.
	Inspired by their benefits in wireless communications, recent studies have also explored MAs for enhancing the performance of wireless sensing \cite{ma2024movable,chen2024exploiting}. Furthermore,  practical issues in MA implementation have been addressed, such as channel estimation \cite{xiao2024channel} and antenna movement delay minimization \cite{li2024minimizing,wang2024throughput}. 
	
	While the existing works on MA mainly focus on the antenna position optimization for wireless channel reconfiguration, there is a growing interest in exploiting antenna orientation/rotation adjustment for performance enhancement. 
	Towards this end, the six-dimensional movable antenna (6DMA) has been recently proposed for flexible adjustment of both the three-dimensional (3D) positions and 3D rotations of distributed antenna arrays \cite{shao20246d,shao20246dd,shao20246dma,shao2024exploiting,shao2024distributed}.
	%equipped with practical directional antennas.
	Compared with position-adjustable MAs, 6DMA introduces some new features. First, 6DMA can more efficiently adapt to wireless channel variations in the 3D space,
	compared to antennas movable only in a specific one-dimensional (1D) line or two-dimensional (2D) region. Second, 6DMA tunes the antenna positions and rotations at the per-array level instead of the per-antenna level for MAs. These new features enable 6DMA to more flexibly and efficiently adapt to 3D large-scale user/target distributions in wireless networks which vary much more slowly than their instantaneous channels/locations, thus leading to a practically appealing implementation with infrequent antenna movement.
	As such, it has been revealed that 6DMA can achieve significant performance gains over FPAs/position-adjustable MAs in both communication  \cite{shao20246d,shao20246dd} and sensing applications \cite{shao2024exploiting}.
	
	Despite the above progress in 6DMA, there remain various challenges in designing and implementing 6DMA that are yet to be addressed.
	First, as 6DMA systems generally need to jointly optimize the 3D positions and 3D rotations of all arrays \cite{shao20246d}, the number of design variables increases significantly with that of the arrays. Thus, it is desirable to reduce the number of position/rotation variables per array so as to reduce the overall optimization complexity.
	Second, the position and rotation variables of each array are both defined in the global Cartesian coordinate system (CCS) in the existing 6DMA model \cite{shao20246d}, rendering their sophisticated coupling in the design objective and practical array movement constraints. 
	As a result, the commonly adopted approach of alternately optimizing the global positions and the global rotations of 6DMA arrays in an iterative manner may get stuck at undesired local optimum. This is mainly due to the fact that when the position of each array is being adjusted, its rotation in the global CCS remains unchanged, and vice versa.
	This makes the array position/rotation adjustments highly confined, since they need to meet practical array movement constraints. 
	Last but not least, although it has been shown that the joint optimization of 6DMA array positions and rotations in the global CCS can significantly improve the performance over traditional FPAs \cite{shao20246d},
	what still remains unknown but is crucial from a system implementation perspective is whether the local rotation adjustment of each array
	at its optimized global position can contribute greatly to the overall performance improvement, which is rather difficult to examine based on the existing 6DMA model/algorithm \cite{shao20246d}.
	
	To address the above issues, we propose a new hierarchically tunable 6DMA (HT-6DMA) model for base station (BS) in this paper.
	Our main results are given as follows:
	\begin{itemize}
		\item The HT-6DMA system is composed of multiple arrays with practical directional antennas that can move freely on a given 3D spherical surface at the BS. Each array is characterized by its position and rotation/orientation. 
		Specifically, the position of each array is	specified by the elevation and azimuth angles of its center with respect to (w.r.t.) the center of the BS in the global spherical coordinate system (SCS), while the orientation is specified by the elevation and azimuth angles of its unit outward normal vector w.r.t. the array's center in the local SCS. The use of global SCS and local SCS for representing the position/orientation of each array is a new feature of the proposed HT-6DMA model, which facilitates the independent adjustments of array positions and rotations in a sequential manner (i.e., hierarchically tunable).
		\item First, we consider an uplink communication scenario, where multiple users each equipped with a single antenna communicate with an HT-6DMA BS in a multiple access channel (MAC). 
		We aim to maximize the average sum rate of the users in the long term via jointly optimizing the (global) positions and (local) rotations of all 6DMA arrays.
		The formulated optimization problem is non-convex and thus difficult to be solved optimally in general. 
		By leveraging the unique structure of HT-6DMA model that  decouples the design of position variables from that of rotation variables, we propose an efficient algorithm for solving this problem, which first designs the arrays' positions and then tunes their local rotations, both via alternating optimization.
		\item Next, we consider the airway sensing scenario, in which an HT-6DMA BS aims to optimize the sensing performance within one or more airway segments in the air. We maximize the minimum received sensing signal power along airway segments by jointly optimizing the (global) positions and (local) rotations of all 6DMA arrays as well as the transmit signal covariance matrix. 
		As the objective function is non-differentiable, we first replace it by a differentiable surrogate function via entropic regularization. Then, similar as the communication scenario, we optimize the positions and rotations of all arrays sequentially, both via alternating optimization under a given transmit covariance matrix.
		After the positions/orientations of all arrays are optimized, the transmit covariance matrix is also optimized  to further enhance the sensing performance. 
		\item Finally, we present numerical results to validate the performance of the proposed algorithms based on the HT-6DMA model in both wireless communication and sensing scenarios. It is shown that the proposed HT-6DMA communication design significantly improves the average sum rate of users compared to that with traditional FPA arrays as well as the existing 6DMA model/algorithm. 
		It is also shown that the proposed HT-6DMA sensing design significantly increases the minimum received sensing signal power over the airway segments compared to that with traditional FPA arrays.
		%		in the airway sensing scenario, where the existing 6DMA algorithm is not viable. 
		Besides, optimizing the transmit covariance matrix leads to more concentrated signal power on the airways,
		thus bringing additional performance gains with the optimized array positions/rotations.
		%		 and thus it is still necessary to design the transmit signaling tailored to the optimized array configuration for better sensing performance.
		In particular, it is unveiled that the communication/sensing performance gains of HT-6DMA mainly come from arrays' position adjustment on the spherical surface, rather than their local orientation adjustment at the optimized positions.
		%		 in both wireless communication and sensing scenarios. 
		Nevertheless, it is also shown that in both scenarios, the additional performance gain from arrays' local orientation adjustment increases when the required minimum distance between arrays increases.
		%		 or the array dimensions become more asymmetric. 
	\end{itemize}
	
	The remainder of the paper is organized as follows.  Section \ref{Section_system_model} presents the system model. Section \ref{sec:Uplink_com} and Section \ref{sec:sensing_problem_formulation} present the proposed algorithms for communication and sensing, respectively. Section \ref{sec:numerical_result} presents numerical results to validate the proposed designs and Section \ref{sec:conclusion} concludes this paper.
	
	{\it Notations:} Boldface letters refer to vectors (lower case) or matrices (upper case). For a matrix $\bm{M}$, $\bm{M}^H$, $\bm{M}^T$, $\bm{M}^{-1}$, $\det (\bm{M})$, $\operatorname{tr} (\bm{M})$ and $\bm{M}\left[:,i\right]$ denote its conjugate transpose, transpose, inverse, determinant, trace, and its $i$-th column, respectively. 
	$\mathbb{R}^{x\times y}$ and $\mathbb{C}^{x\times y}$ represent the spaces of real and complex matrices with dimension $x \times y$, respectively. {${\mathbb{E}}\{\cdot\}$} denotes the statistical expectation. $\|\bm{x}\|$ is the Euclidean norm of a vector $\bm{x}$.
	$\mathrm{j} = \sqrt{-1}$ is the imaginary unit. The distribution of a circularly symmetric complex Gaussian (CSCG) random vector with mean zero and covariance matrix $\bm{\Sigma}$ is denoted by $\mathcal{CN}(\bm{0},\bm{\Sigma})$, and $\sim$ stands for “distributed as”. $\min \left\{a,b\right\}$ denotes the operation that returns the minimum of the two values $a$ and $b$. $\lfloor \cdot \rfloor$ denotes the flooring operation.

	\section{System Model}\label{Section_system_model}
	
	This section presents the HT-6DMA BS model,
	followed by the corresponding channel model and performance metrics for both wireless communication and sensing.
	
	\subsection{HT-6DMA BS Model}\label{sec:System_SMA}
	
\begin{figure}[t]
		\centering
		\includegraphics[width=2.3in]{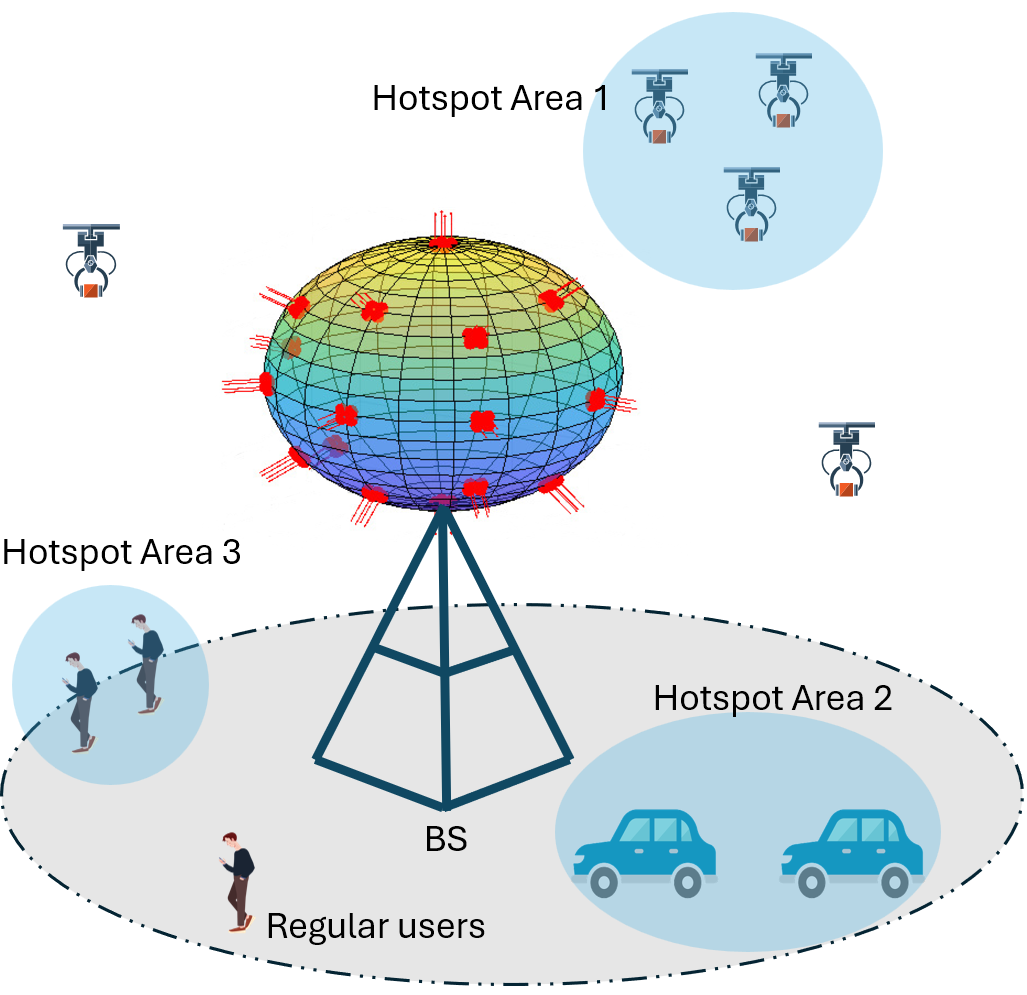}
		\centering
		\caption{An HT-6DMA BS with $B=16$ UPAs and $N=4$ antennas/UPA.}
		\label{fig:SMA_inst}
	\end{figure}

	The HT-6DMA BS is equipped with $B$ uniform planar arrays (UPAs), denoted as $\mathcal{B} = \{1,2,...,B\}$, and each array consists of $N \geq 1$ antennas, denoted by $\mathcal{N} = \{1, 2, ..., N\}$. As shown in Fig. \ref{fig:SMA_inst}, all the arrays are movable on the surface of a sphere with radius $R$\footnote{Here we assume that the distance between the array's center to the BS's center is the same for all the arrays, while it can also be tunable for each individual array for more design flexibility \cite{shao20246d}, which is left for future work.}, which is denoted as $\mathcal{C}$. 
	Each array's position on the surface of $\mathcal{C}$ is characterized by the elevation angle and azimuth angle of its center w.r.t. the origin of the global SCS (assumed to be the center of the BS, i.e., the center of $\mathcal{C}$), which is denoted by
	\begin{align}
		\bm{t}_b &= \left[ \theta_b, \phi_b \right]^T, \enspace  b \in \mathcal{B},
	\end{align}
	where $\theta_b \in  \left[-\frac{\pi}{2},\frac{\pi}{2}\right]$ and $\phi_b \in \left[-\pi,\pi\right]$. In addition, the rotation of each array is characterized by the elevation angle and azimuth angle of the array's unit outward normal vector w.r.t. the origin of its local SCS (assumed to be the center of the array), denoted by
	\begin{align}
		\bm{u}_b = \left[ \vartheta_b, \varphi_b \right]^T, \enspace  b \in \mathcal{B},
	\end{align}
	where $\vartheta_b \in \left[0,\frac{\pi}{2}\right]$ and $\varphi_b \in \left[-\pi,\pi\right]$. 
	Let $\bm{r}_{b,n}^l \in \mathbb{R}^{3 \times 1}$ denote the original position of the $n$-th antenna of the $b$-th array in its local CCS and $\bm{r}_{b,n}^g \in \mathbb{R}^{3 \times 1}$ denote its corresponding transformed position in the global CCS, respectively. Under given $\bm{t}_b$ and $\bm{u}_b$, the relation between $\bm{r}_{b,n}^g$ and $\bm{r}_{b,n}^l$ is given by
	\begin{align}\label{eq:coord_local2global}
		\bm{r}_{b,n}^g = R \bm{l}_b(\bm{t}_b) + \bm{R}(\bm{t}_b) \bm{R}(\bm{u}_b) \bm{r}_{b,n}^l, \enspace n \in \mathcal{N}, \enspace b \in \mathcal{B},
	\end{align}
	where $\bm{R}(\bm{u}_b)$
	denotes the matrix that transforms the original local coordinates of antenna positions of the $b$-th array into their corresponding rotated local coordinates specified by $\bm{u}_b$, and $\bm{R}(\bm{t}_b)$ denotes the matrix that further
	transforms the rotated local coordinates of antenna positions of the $b$-th array into their corresponding intermediate global coordinates specified by $\bm{t}_b$. To get the final global coordinates $\bm{r}_{b,n}^g$, the distance shift $R \bm{l}_b(\bm{t}_b)$ needs to be added in (\ref{eq:coord_local2global}), 
	%	which is essentially the position of the $b$-th array on the sphere $\mathcal{C}$ with radius $R$ in the global CCS, 
	where $\bm{l}_b(\bm{t}_b)$ is given by
	\begin{align}\label{eq:l_tb_relation}
		\bm{l}_b(\bm{t}_b) = \left[\cos\theta_b \cos\phi_b, \cos\theta_b \sin\phi_b, \sin\theta_b\right]^T.
	\end{align}
	To derive $\bm{R}(\bm{u}_b)$ and $\bm{R}(\bm{t}_b)$ in (\ref{eq:coord_local2global}), 	
	since the rotation matrices w.r.t. $\mr{y}$-axis and $\mr{z}$-axis are given by
	\begin{align}
		\nonumber
		\bm{R}_\mr{y}(y)   = \left[
		\arraycolsep=2pt\def\arraystretch{1.2}
		\begin{array}{ccc}
			\cos y  & 0 & \sin y  \\[1pt]
			0 &  1 & 0 \\[1pt]
			-\sin y & 0 & \cos y  \\
		\end{array}\right], \bm{R}_\mr{z}(z) = \left[
		\arraycolsep=2pt\def\arraystretch{1.2}
		\begin{array}{ccc}
			\cos z & -\sin z & 0 \\[1pt]
			\sin z & \cos z & 0 \\[1pt]
			0 & 0 & 1 \\
		\end{array}\right],
	\end{align}
	respectively, $\bm{R}(\bm{u}_b)$ is expressed as
	\begin{align}\label{eq:R_ub_def}
		\bm{R}(\bm{u}_b) & = \bm{R}_\mr{z}({\varphi_b }) \bm{R}_\mr{y}({\frac{\pi}{2}-\vartheta_b }),
	\end{align}
	and $\bm{R}(\bm{t}_b)$ is expressed as
	\begin{align}\label{eq:R_tb_def}
		\bm{R}(\bm{t}_b ) =  \bm{R}_\mr{z}({\phi_b }) \bm{R}_\mr{y}({\frac{\pi}{2}-\theta_b }).
	\end{align}	
	
\begin{figure}[t]
		\centering
		\includegraphics[width=3.2in]{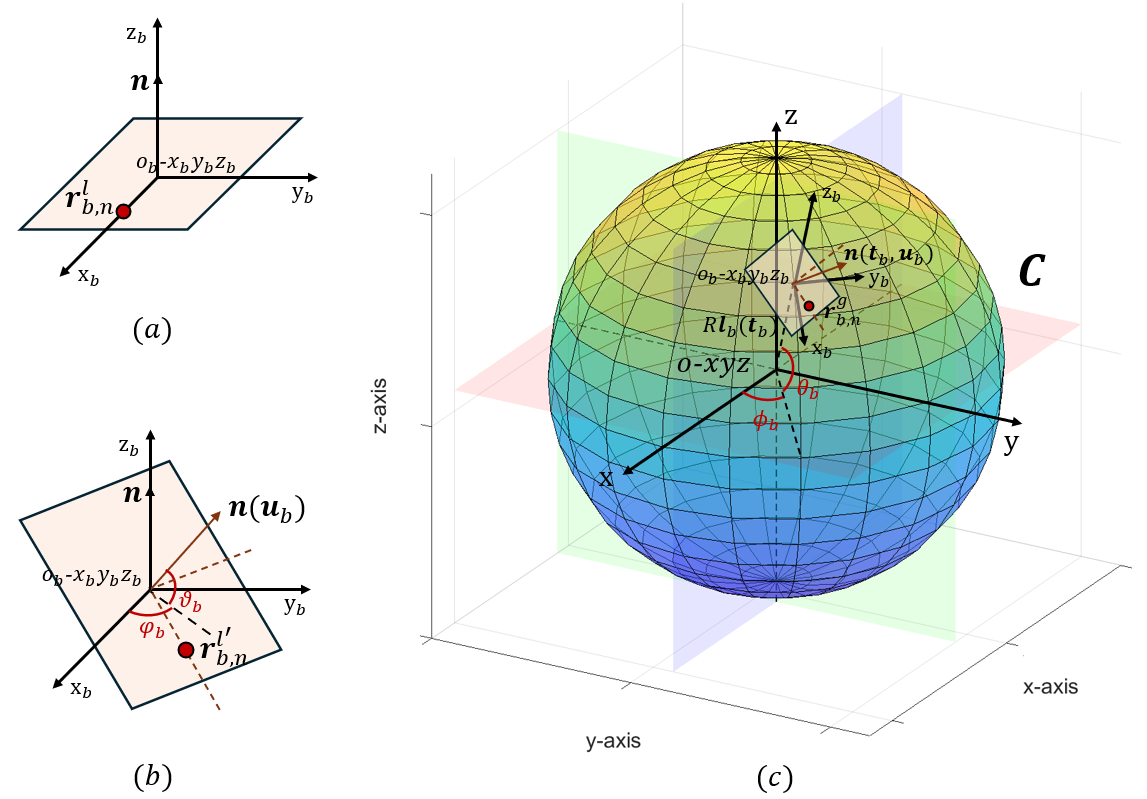}
		\centering
		\caption{The transformation of the local coordinates of antenna positions $\bm{r}_{b,n}^l$ to their corresponding global coordinates $\bm{r}_{b,n}^g$ in (\ref{eq:coord_local2global}).}
		\label{fig:spherical_region_intermediate}
	\end{figure}

	Fig. \ref{fig:spherical_region_intermediate} shows the transformation procedure in (\ref{eq:coord_local2global}) and how we get $\bm{R}(\bm{u}_b)$ and $\bm{R}(\bm{t}_b)$ in (\ref{eq:R_ub_def}) and (\ref{eq:R_tb_def}), respectively. 
	Before the local rotation, the original local unit outward normal vector $\bm{n} = \left[0,0,1\right]^T$ is assumed to
	be aligned with the local $\mr{z}_b$-axis, as shown in Fig. \ref{fig:spherical_region_intermediate}(a). After the local rotation, $\bm{n}$ is rotated towards $\bm{n}(\bm{u}_b)$ according to $\bm{u}_b = \left[\vartheta_b,\varphi_b\right]^T$, where $\bm{n}(\bm{u}_b)$ denotes the new local unit outward normal vector of the array after the local rotation. The overall local rotation is composed of two separable rotations, where the first one is to rotate $\bm{r}_{b,n}^l$ around the local $\mr{y}_b$-axis by $\frac{\pi}{2} - \vartheta_b$, followed by the second rotation around the local $\mr{z}_b$-axis by $\varphi_b$, which leads to the expression of $\bm{R}(\bm{u}_b)$ in (\ref{eq:R_ub_def}).
	Acccordingly,
	after the local rotation specified by $\bm{R}(\bm{u}_b)$, 
	the original local coordinate of the $n$-th antenna of the $b$-the array $\bm{r}_{b,n}^l$ in Fig. \ref{fig:spherical_region_intermediate}(a) is transformed into its rotated local coordinate in Fig. \ref{fig:spherical_region_intermediate}(b), which is denoted as $\bm{r}_{b,n}^{l'}$ and is given by
	\begin{align}\label{eq:local_intermediate}
		\bm{r}_{b,n}^{l'} = \bm{R}(\bm{u}_b) \bm{r}_{b,n}^l, \enspace n \in \mathcal{N}.
	\end{align} 
	Then, to transform the obtained $\bm{r}^{l'}_{b,n}$ in the corresponding local CCS $o_b$-$x_by_bz_b$ into $\bm{r}^{g}_{b,n}$ in the global CCS $o$-$xyz$ in (\ref{eq:coord_local2global}), we further rotate $\bm{r}^{l'}_{b,n}$ around the global $\mr{y}$-axis by $\frac{\pi}{2} - \theta_b$, followed by the rotation around the global $\mr{z}$-axis by $\phi_b$, as shown in Fig. \ref{fig:spherical_region_intermediate}(c). This leads to the expression of  $\bm{R}(\bm{t}_b)$ in (\ref{eq:R_tb_def}). Besides, $\bm{n}(\bm{u}_b)$ can be further denoted as $\bm{n}(\bm{t}_b,\bm{u}_b)$ in the global CCS, as shown in Fig. \ref{fig:spherical_region_intermediate}(c).  After the global rotation specified by $\bm{R}(\bm{t}_b)$, the distance shift $R \bm{l}_b(\bm{t}_b)$ is added to yield $\bm{r}^{g}_{b,n}$ in (\ref{eq:coord_local2global}).
	
	%	and finally get the local CS $o_b$-$x_by_bz_b$ shown in Fig. \ref{fig:spherical_region_intermediate}. The above procedure leads to (\ref{eq:spherical_MA_key_expression}). The derivation of (\ref{eq:R_ub_def}) can be obtained similarly, but w.r.t. the local CS $o_b$-$x_by_bz_b$, as also shown in Fig. \ref{fig:spherical_region_intermediate}. Combing both local
	%	and global coordinate transformations with the distance shift $R \bm{l}_b(\bm{t}_b)$ leads to (3).
	
	\begin{remark}
		\emph{We provide an example to demonstrate the differences between the proposed HT-6DMA model and the existing 6DMA model in \cite{shao20246d}. Without loss of generality, we set $R = 1$ for the sphere and consider an array on the spherical surface where its center is located at $\left[ \frac{1}{\sqrt{3}}, \frac{1}{\sqrt{3}}, \frac{1}{\sqrt{3}} \right]$ in the global CCS without any local rotation. In the  HT-6DMA model, the array's position and local rotation is characterized by $\left[\theta_b, \phi_b\right] = \left[\frac{\pi}{4}, \frac{\pi}{4}\right]$ and $\left[\vartheta_b, \varphi_b\right] = \left[\frac{\pi}{2}, 0\right]$, respectively. In contrast, the same array is characterized by position vector $\left[x_b, y_b, z_b\right] = \left[ \frac{1}{\sqrt{3}}, \frac{1}{\sqrt{3}}, \frac{1}{\sqrt{3}} \right]$ and rotation vector $\left[\alpha_b, \beta_b, \gamma_b\right] = \left[ -\frac{\pi}{6}, \arcsin(\frac{1}{\sqrt{3}}), -\arcsin(\frac{\sqrt{6}}{3})\right]$ in \cite{shao20246d}. 
			It is observed that the HT-6DMA model not only needs fewer parameters (4 instead of 6) to characterize the array position/rotation, but also admits a simpler representation of rotation w.r.t. the local SCS compared with that in \cite{shao20246d} w.r.t. the global CCS.}
	\end{remark}
	
	Next, we consider practical constraints in the proposed HT-6DMA model on antenna position/orientation adjustment, which are according to those given in \cite{shao20246d} but now take different forms based on our new model as follows.
	\begin{enumerate}
		\item \textit{Minimum distance constraints:} A minimum distance requirement between any pair of arrays has to be imposed on the HT-6DMA model to avoid overlapping between arrays, which is expressed as
		\begin{align}\label{eq:dist_constr}
			R \|\bm{l}_i(\bm{t}_i) - \bm{l}_j(\bm{t}_j)\| \geq d_{\text{min}}, \enspace \forall i,j \in \mathcal{B}, j \neq i,
		\end{align}
		where $d_{\text{min}}$ is the given minimum distance required between the centers of any two arrays.
		\item \textit{Rotation constraints to avoid signal reflection:} The arrays must meet the rotation constraints to avoid mutual signal reflections between any two arrays, i.e.,
		\begin{align}\label{eq:constr_signal_reflection}
			\bm{n}(\bm{t}_i,\bm{u}_i)^T (\bm{l}_j(\bm{t}_j) - \bm{l}_i(\bm{t}_i)) \leq 0, \enspace \forall i,j \in \mathcal{B}, j \neq i,
		\end{align}
		where $\bm{n}(\bm{t}_i,\bm{u}_i)$ is the global unit outward normal vector of the $i$-th array. % notice that $\bm{n}(\bm{t}_i,\bm{u}_i)$ is just $\bm{n}(\bm{t}_i)$ shown in Fig. 2(c), but viewed from the global CCS.
		Based on (\ref{eq:coord_local2global}), $\bm{n}(\bm{t}_i,\bm{u}_i)$ is expressed as
		\begin{align}\label{eq:constr_signal_reflection_loc}
			\bm{n}(\bm{t}_i,\bm{u}_i) = \bm{R}(\bm{t}_i) \bm{R}(\bm{u}_i) \bm{n},
		\end{align}
		where $\bm{n} = \left[0,0,1\right]^T$ is the aforementioned original local unit outward normal vector and is the same for all arrays, as shown in Fig. \ref{fig:spherical_region_intermediate}(a). 
		As such, the constraint in (\ref{eq:constr_signal_reflection}) is then simplified as
		\begin{align}
			\nonumber
			(\bm{R}(\bm{u}_i)\left[:,3\right])^T \bm{R}(\bm{t}_i)^T & (\bm{l}_j(\bm{t}_j) - \bm{l}_i(\bm{t}_i)) \leq 0, \\
			\label{eq:reflection_constr}
			& \forall i,j \in \mathcal{B}, j \neq i.
		\end{align}
		\item \textit{Rotation constraints to avoid signal blockage:} As each array shall not rotate towards the center of the BS which leads to signal blockage, we have
		\begin{align}\label{eq:blockage_constr}
			\bm{n}(\bm{t}_b,\bm{u}_b)^T \bm{l}_b(\bm{t}_b) \geq 0, \enspace \forall b \in \mathcal{B}.
		\end{align}
		This constraint can be shown to be already satisfied in our new model since $\vartheta_b \in \left[0, \frac{\pi}{2}\right]$.
		Thus, it is omitted in the sequel of this paper.
	\end{enumerate}
	
	From (\ref{eq:coord_local2global}), it is worth pointing out that in the proposed HT-6DMA model, the array positions and rotations are hierarchically tunable in the sense that each array's position on the sphere, i.e., $\bm{t}_b$, can be adjusted first subject to the practical positioning constraint in (\ref{eq:dist_constr}) without any local rotation, followed by the design of the local rotation of each array, i.e., $\bm{u}_b$, subject to the practical rotation constraint in (\ref{eq:reflection_constr}). As will be shown later in Section \ref{sec:numerical_result}, this hierarchically tunable structure can greatly simplify the algorithm design for the array position/rotation joint optimization, while achieving a better performance than the existing 6DMA design \cite{shao20246d}.

	\subsection{Uplink Communication}

	In this subsection, we consider the uplink communication scenario where $K$ users located in the far-field region of the HT-6DMA BS communicate with the BS simultaneously. Each user is equipped with a single omni-directional FPA.
	For the ease of illustration, let $\bm{t} = \left[ \bm{t}_1, \bm{t}_2, ..., \bm{t}_B \right] \in \mathbb{R}^{2 \times B}$ and $\bm{u} = \left[ \bm{u}_1, \bm{u}_2, ..., \bm{u}_B \right] \in \mathbb{R}^{2 \times B}$ denote the stacked position variables and rotation variables for all $B$ arrays at the BS, respectively.

	\subsubsection{3D Steering Vector}  Let $\mathcal{K} \in \{1,2,...,K\}$ denote the set of users, and $\omega_k \in \left[-\frac{\pi}{2}, \frac{\pi}{2}\right]$ and $\psi_k \in \left[-\pi, \pi\right]$ denote the elevation angle and the azimuth angle of the location of the $k$-th user w.r.t. the center of the BS, $k \in \mathcal{K}$. % no need to specify w.r.t. the global CCS or global SCS here, SCS is defined also based on x,y,z axis.
	The coordinates of its unit-length pointing vector in the global CCS is
	given by
	\begin{align}\label{eq:point_vec}
		\bm{f}_k = \left[\cos \omega_k \cos \psi_k, \cos\omega_k \sin \psi_k, \sin\omega_k\right]^T.
	\end{align}
	Based on (\ref{eq:coord_local2global}) and (\ref{eq:point_vec}), 
	the steering vector of the $b$-th array corresponding to the $k$-th user is given by
	\begin{align} 
		\bm{a}_k(\bm{t}_b,\bm{u}_b) = \left[e^{-\mr{j}\frac{2\pi}{\lambda}\bm{f}_k^T \bm{r}_{b,1}^g},...,e^{-\mr{j}\frac{2\pi}{\lambda}\bm{f}_k^T \bm{r}_{b,N}^g}\right]^T,  \forall b \in \mathcal{B}, 
	\end{align} 
	where $\lambda$ denotes the carrier wavelength.
	
	\subsubsection{Effective Antenna Gain} Let the elevation angle and the azimuth angle of the location of the $k$-th user w.r.t. the center of the $b$-th array in its local SCS be denoted by $\tilde{\omega}_{bk} = \arcsin (\tilde{z}_{bk})$ and $\tilde{\psi}_{bk} = \operatorname{arctan2}\left(\tilde{y}_{bk}, \tilde{x}_{bk}\right)$, respectively, where $\left[\tilde{x}_{bk},\tilde{y}_{bk},\tilde{z}_{bk}\right]^T = (\bm{R}(\bm{t}_b) \bm{R}(\bm{u}_b))^{-1} \bm{f}_k$ denotes the coordinates of the unit-length pointing vector in the local CCS of the $b$-th array.
	The effective antenna gain of the $k$-th user w.r.t. the $b$-th array in the linear scale is then given by
	\begin{align}\label{eq:directional_gain}
		g_k(\bm{t}_b,\bm{u}_b) = 10^{\frac{A(\tilde{\omega}_{bk},\tilde{\psi}_{bk})}{10}},
	\end{align}
	where $A(\tilde{\omega}_{bk},\tilde{\psi}_{bk})$ specifies the directional gain of each antenna in terms of 
	$\tilde{\omega}_{bk}$ and $\tilde{\psi}_{bk}$ \cite{3gpp2018study}. 
	
	\subsubsection{Effective Channel} For the ease of exposition, we assume the line-of-sight (LoS) channel between each user and the BS. The channel between the $k$-th user and the BS is then expressed as
	\begin{align}
		\bm{h}_k(\bm{t},\bm{u}) = \sqrt{\nu_k} \left[\mathcal{A}_k(\bm{t}_1,\bm{u}_1),...,\mathcal{A}_k(\bm{t}_B,\bm{u}_B)\right]^T,
	\end{align}
	where
	\begin{align}
		\mathcal{A}_k(\bm{t}_b,\bm{u}_b) = \sqrt{g_k(\bm{t}_b,\bm{u}_b)}\bm{a}_k(\bm{t}_b,\bm{u}_b), \enspace \forall b \in \mathcal{B},
	\end{align}
	and $\nu_k = \epsilon_0 d_k^{-\zeta_c}$ is the path gain, where $\epsilon_0$ is a constant,
	%	related to reference channel power, 
	$d_k$ is the distance between the $k$-th user and the BS, and $\zeta_c$ is the path-loss exponent. % you start from here.
	The received signals at the BS from all the $K$ users are thus given by
	\begin{align}
		\bm{y} = \bm{H}(\bm{t},\bm{u}) \bm{x} + \bm{z},
	\end{align}
	where $\bm{x} = \sqrt{p}\left[x_1,x_2,...,x_K\right]^T$ and the transmit power $p$ is assumed to be the same for all the users. $\bm{H}(\bm{t},\bm{u}) = \left[\bm{h}_1(\bm{t},\bm{u}),\bm{h}_2(\bm{t},\bm{u}),...,\bm{h}_K(\bm{t},\bm{u})\right] \in \mathbb{C}^{NB \times K}$. $\bm{z} \sim \mathcal{CN}(\bm{0}_{NB}, \sigma^2 \bm{I}_{NB})$ denotes the CSCG noise random vector.
	%	with zero mean and covariance matrix $\sigma^2 \bm{I}_{NB}$.
	With perfect channel state information (CSI) at the BS, Gaussian signaling, and multiuser joint decoding with the MMSE (minimum mean square error)-SIC (successive interference cancellation) receivers \cite{Tse2005book}, the maximum achievable sum rate for all the users in their
	uplink communication is 
	\begin{align}
		C(\bm{t},\bm{u}) = \log_2 \det \left(\bm{I}_{NB} + \frac{p}{\sigma^2} \bm{H}(\bm{t},\bm{u}) \bm{H}(\bm{t},\bm{u})^H\right).
	\end{align} 
	By applying the expectation w.r.t. $\bm{H}(\bm{t},\bm{u})$ due to random user locations, we obtain the average sum rate of users as
	\begin{align}\label{eq:exp_cap}
		C_{\text{avg}} = \mathbb{E}_{\bm{H}} \left[C(\bm{t},\bm{u})\right].
	\end{align}
	Since it is difficult to derive (\ref{eq:exp_cap}) analytically, we approximate it by using the standard Monte Carlo (MC) method \cite{shao20246d}, 
	and the average sum rate is approximated by
	\begin{align}\label{eq:mc_cap}
		\hat{C}(\bm{t},\bm{u}) = \frac{1}{S} \sum_{s=1}^{S} C_s(\bm{t},\bm{u}),
	\end{align}
	where $C_s(\bm{t},\bm{u}) = C(\bm{t},\bm{u})| \bm{H}_s$ denotes the sum rate of the $s$-th realization given the channel sample $\bm{H}_s$ and $S$ is the total number of channel samples based on the distribution of $\bm{H}$ or user spatial distribution, which will be specified in Section \ref{sec:numerical_result}.
	
	\subsection{Airway Sensing}
	
	In this subsection, we consider the airway sensing scenario where the HT-6DMA BS aims to enhance the sensing performance over airways located in its far-field region.
	As in Section \ref{sec:System_SMA}, the HT-6DMA BS is centered at $\left[0,0,0\right]^T$ in the global CCS. There is a straight airway segment above the BS with its starting point at $\bm{r}_s = \left[x_s,y_s,z_s\right]^T$ and end point at $\bm{r}_e = \left[x_e,y_e,z_e\right]^T$\footnote{We assume that the other parts of the airway are covered by other BSs. In addition, we focus on the case of single airway segment in this section while our results can be extended to multiple airway line segments, as shown in Section \ref{sec:sensing_numerial}.}. The corresponding elevation angle and azimuth angle of these two points w.r.t. the center of BS are thus given by
	\begin{align}
		\nonumber
		\theta_s & = \operatorname{arctan2} (z_s,\sqrt{x_s^2 + y_s^2}), \enspace \theta_e = \operatorname{arctan2} (z_e, \sqrt{x_e^2 + y_e^2}), \\
		\phi_s & = \operatorname{arctan2} (y_s,x_s), \enspace \phi_e = \operatorname{arctan2} (y_e,x_e).
	\end{align}
	\begin{figure}[t] 
		\centering
		\setlength{\abovecaptionskip}{+4mm}
		\setlength{\belowcaptionskip}{+1mm}
		\subfigure[Airway $\bm{r}(\xi)$ in the 3D CCS.]{ \label{fig:Airway_seg_3D}
			\includegraphics[width=1.672in]{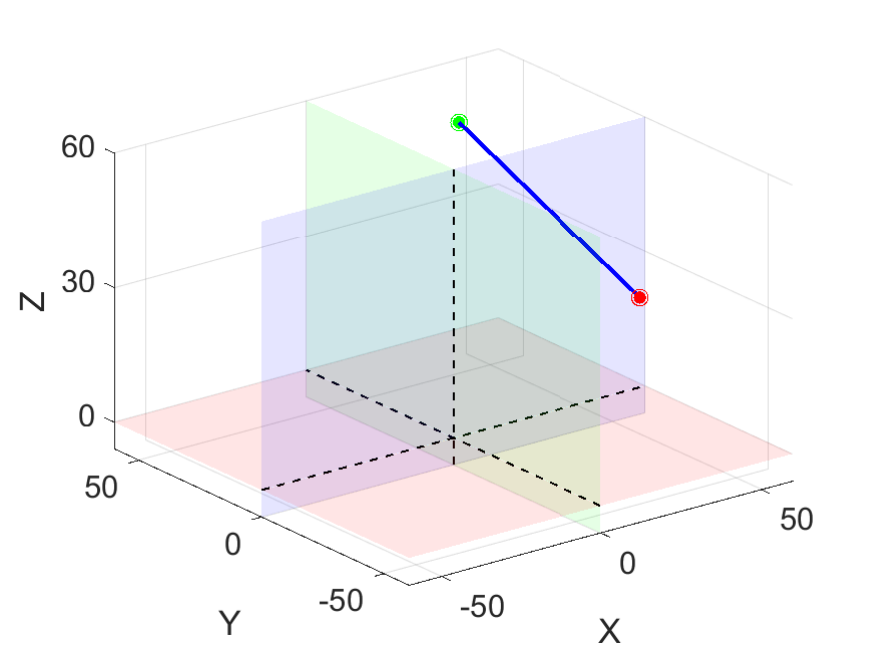}} % 1.672/2.8
		\subfigure[Airway $\bm{r}_p(\xi)$ in the 2D angular plot.]{ \label{fig:Airway_seg_polar}
			\includegraphics[width=1.672in]{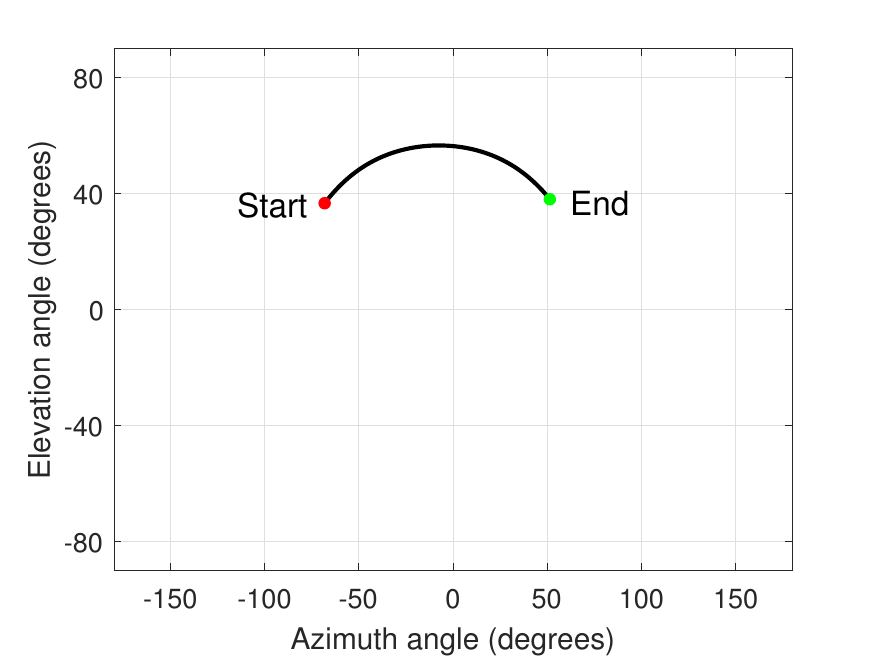}}
		\caption{Illustration of the airway segment.}
		\label{fig:airway_example}
	\end{figure}
	Each point on the airway segment is then expressed as
	\begin{align}
		\nonumber
		\bm{r}(\xi) = (1-\xi)\bm{r}_s + \xi \bm{r}_e = \left[x(\xi),y(\xi),z(\xi)\right]^T, \xi \in [0,1],
	\end{align}
	where $v(\xi) = (1-\xi)v_s + \xi v_e, \enspace v \in \{x,y,z\}.$
	As the airway is located in the far-field region of the BS, given a specific point $\bm{r}(\xi)$ on the airway segment, the corresponding steering vector of each 6DMA array is only related to its angle direction w.r.t. the BS center. 
	Thus, we consider the more concise 2D angular representation of each point $\bm{r}(\xi)$, which is represented by
	\begin{align}
		\bm{r}_p (\xi)  = \left[\theta(\xi),\phi(\xi)\right]^T, 
	\end{align}
	where $\theta(\xi) = \operatorname{arctan2} (z(\xi),\sqrt{x(\xi)^2+ y(\xi)^2})$ and $\phi(\xi) = \operatorname{arctan2} (y(\xi),x(\xi)).$
	An example of the airway segment is shown in Fig. \ref{fig:airway_example} for illustration.
	As the pointing vector $\bm{f}_\xi$ for the direction $(\theta(\xi),\phi(\xi))$ is given as
	\begin{align}
		\bm{f}_\xi = \left[\cos\theta(\xi)\cos\phi(\xi),\cos\theta(\xi)\sin\phi(\xi), \sin\theta(\xi)\right]^T,
	\end{align}
	the corresponding steering vector of the $b$-th array at the BS is given by
	\begin{align}
		\bm{a}_\xi(\bm{t}_b,\bm{u}_b) = \left[e^{-\mr{j}\frac{2\pi}\lambda\bm{f}_\xi^T \bm{r}_{b,1}^g},...,e^{-\mr{j}\frac{2\pi}\lambda\bm{f}_\xi^T \bm{r}_{b,N}^g}\right]^T, \forall b \in \mathcal{B}.
	\end{align}
	Let $\bm{x}_r \in \mathbb{C}^{NB \times 1}$ denote the transmit sensing signal with zero mean and covariance matrix $\bm{R}_d$ of full rank, i.e.,
	\begin{align}
		\bm{R}_d = \mathbb{E}\left[\bm{x}_r \bm{x}_r^H\right].
	\end{align}
	Suppose that the maximum transmit power budget at the BS is $P_0$. Then, the transmit sensing signal $\bm{x}_r$ needs to satisfy the average sum-power constraint given as
	\begin{align}\label{eq:sum-power}
		\mathbb{E}\left[\|\bm{x}_r\|^2\right] = \operatorname{tr}(\bm{R}_d) \leq P_0.
	\end{align}
	As airways are typically located far above the ground and can be pre-selected to avoid signal blockage, we asssume LoS channels between the BS and all points on the airway segment. 
	To enhance the performance of various sensing tasks such as unmanned aerial vehicles (UAVs)/drones' detection and parameter estimation \cite{hua2023optimal} along the airway, the received sensing signal power distribution along the airway is adopted as the airway sensing performance metric, which is given by
	\begin{align}\label{eq:trans_power_beampattern}
		\mathcal{P}(\theta(\xi),\phi(\xi)) =\mathbb{E} \left[|\bm{h}_\xi^T(\bm{t},\bm{u}) \bm{x}_r|^2\right]=
		\bm{h}_\xi^T(\bm{t},\bm{u}) \bm{R}_d \bm{h}_\xi^*(\bm{t},\bm{u}),
	\end{align}
	where $\bm{h}_\xi(\bm{t},\bm{u})$ denotes the sensing channel between a point $\bm{r}_p(\xi)$ on the airway and the BS and is given by
	\begin{align}
		\bm{h}_\xi(\bm{t},\bm{u}) = \sqrt{\nu(\xi)} \left[\mathcal{A}_\xi(\bm{t}_1,\bm{u}_1),...,\mathcal{A}_\xi(\bm{t}_B,\bm{u}_B)\right]^T
	\end{align}
	with
	\begin{align}
		\mathcal{A}_\xi(\bm{t}_b,\bm{u}_b) = \sqrt{g_\xi(\bm{t}_b,\bm{u}_b)}\bm{a}_\xi(\bm{t}_b,\bm{u}_b), \enspace \forall b \in \mathcal{B},
	\end{align}
	and
	\begin{align}
		\nu(\xi) = \epsilon_0 d^{-\zeta_s}_\xi =  \epsilon_0 (\sqrt{x^2(\xi)+y^2(\xi)+z^2(\xi)})^{-\zeta_s},
	\end{align}
	where $\zeta_s$ is the path-loss exponent in the airway sensing scenario.
	Here, $\nu(\xi)$ is assumed to be equal for all 6DMA arrays at the BS since the (shortest) distance between the BS and the airway is practically much larger than the 6DMA moving region's sphere radius $R$.

	\section{Array Position and Rotation Optimization for Uplink Multiuser Communication}\label{sec:Uplink_com}
	
	For the uplink communication scenario, we aim to maximize (\ref{eq:mc_cap}) by jointly optimizing the positions and rotations of all arrays under their practical movement constraints in (\ref{eq:dist_constr}) and (\ref{eq:constr_signal_reflection}).
	The optimization problem
	% based on the SMA system
	is thus formulated as
	\begin{subequations}
		\begin{align}
			(\text{P}1): & \max_{\bm{t},\bm{u}} \hat{C}(\bm{t},\bm{u}) \\
			\text{s.t. } 
			\label{eq:constr_reflection}
			%			& \left[s_{\beta_i}, -s_{\alpha_i}c_{\beta_i}, c_{\alpha_i} c_{\beta_i} \right]  \bm{R}(\bm{a}_i)^T (\bm{l}_j - \bm{l}_i) \leq 0, \\
			& (\bm{R}(\bm{u}_i)\left[:,3\right])^T  \bm{R}(\bm{t}_i)^T (\bm{l}_j(\bm{t}_j) - \bm{l}_i(\bm{t}_i)) \leq 0, \\
			\label{eq:constr_dist_SMA}
			& \|\bm{l}_i(\bm{t}_i) - \bm{l}_j(\bm{t}_j)\| \geq d_{\text{min}}/R, \enspace \forall i,j \in \mathcal{B}, j \neq i.
		\end{align}
	\end{subequations}

	It is worth discussing the differences between (P1) and that based on the existing 6DMA model given in \cite{shao20246d}.   
	First and foremost, the rotation variables  $\bm{u}$ in the HT-6DMA model are defined w.r.t. the local SCS of each array, while those in \cite{shao20246d} are defined in the global CCS. This not only removes the signal blockage constraint in (\ref{eq:blockage_constr}), but also \textit{decouples} the optimization of position and rotation such that the position variables $\bm{t}$ can be designed first subject to the positioning constraint (\ref{eq:constr_dist_SMA}) only, followed by designing the local rotation variables $\bm{u}$ subject to the rotation constraint (\ref{eq:constr_reflection}) only. 
	Second, both the position variables $\bm{t}$ and the local rotation variables $\bm{u}$ in (P1) are defined in the SCS instead of those in the CCS in \cite{shao20246d}, thus reducing the number of variables per array (from \textit{6} to \textit{4}). 
	Third, the moving region $\mathcal{C}$ is a spherical surface for HT-6DMA versus a cube in \cite{shao20246d}. Although $\mathcal{C}$ is a non-convex set for HT-6DMA, we can exploit its property that the distances between all the array centers and the BS center are identical to simplify the algorithm for solving (P1).
	The above differences pave the way for devising a simpler and more efficient algorithm for solving (P1), detailed in the following.

	\subsection{Optimization of $\bm{t}$ with Fixed $\bm{u}$}\label{sec:P1-t_opt_t} 
	
	By exploiting the HT-6DMA model, we solve (P1) by first optimizing $\bm{t}$ with $\bm{u}$ fixed as $\bm{u}_i = \left[\pi/2,0\right]^T, \forall i \in \mathcal{B}$.
	In this case, on the left hand side (LHS) of constraint (\ref{eq:constr_reflection}), we obtain 
	\begin{align}
		\bm{R}(\bm{t}_i) (\bm{R}(\bm{u}_i)\left[:,3\right]) = \bm{R}(\bm{t}_i) \left[0,0,1\right]^T = \bm{l}_i(\bm{t}_i), \forall i \in \mathcal{B}.
	\end{align}
	Then, based on the Cauchy-Schwarz inequality, it follows that
	\begin{align}
		\bm{l}_i^T(\bm{t}_i) (\bm{l}_j(\bm{t}_j) - \bm{l}_i(\bm{t}_i)) \leq 0, \enspace \forall i, j \in \mathcal{B}, j \neq i. 
	\end{align}
	Thus, constraint (\ref{eq:constr_reflection}) is always met and can be removed for optimizing $\bm{t}$. The optimization of $\bm{t}$ in (P1) is thus simplified to
	\begin{subequations}
		\begin{align}
			(\text{P1-t}): & \max_{\bm{t}} \hat{C}(\bm{t}) \\
			\text{s.t. } 
			\label{eq:constr_dist_SMA_ss}
			& \|\bm{l}_i(\bm{t}_i) - \bm{l}_j(\bm{t}_j)\| \geq d_{\text{min}}/R, \enspace \forall i,j \in \mathcal{B}, j \neq i.
		\end{align}
	\end{subequations}
	
	We solve (P1-t) via alternating optimization. Specifically, we optimize each $\bm{t}_b$ with given $\{\bm{t}_j\}_{j \in \mathcal{B} \setminus b}$ in each iteration. 
	Accordingly, (P1-t) in each iteration is expressed as
	\begin{subequations}
		\begin{align}
			(\text{P1-t1}): & \max_{\bm{t}_b} \hat{C}(\bm{t}_b) \\
			\text{s.t. } 
			\label{eq:constr_qb_dist}
			& \|\bm{l}_b(\bm{t}_b) - \bm{l}_j(\bm{t}_j)\| \geq d_{\text{min}}/R, \enspace \forall j \in \mathcal{B} \setminus b.
		\end{align}
	\end{subequations} 
	As (\ref{eq:constr_qb_dist}) is imposed on $\bm{l}_b$ and there exists a one-to-one mapping between $\bm{t}_b$ and its corresponding $\bm{l}_b$, we choose to optimize $\{\bm{l}_b\}$ for convenience. (P1-t1) is thus equivalent to
	\begin{subequations}
		\begin{align}
			(\text{P1-t2}): & \max_{\bm{l}_b} \hat{C}(\bm{l}_b) \\
			\text{s.t. } 
			\label{eq:constr_qb_dist_lb}
			& \|\bm{l}_b - \bm{l}_j\| \geq d_{\text{min}}/R, \enspace \forall j \in \mathcal{B} \setminus b,\\
			\label{eq:constr_unit_lb}
			& \|\bm{l}_b\| = 1, % but this is non-convex..
		\end{align}
	\end{subequations}
	where (\ref{eq:constr_unit_lb}) is to ensure that $\bm{l}_b$ has a unit length.
	(P1-t2) is still difficult to solve due to the non-convex constraints (\ref{eq:constr_qb_dist_lb}) and (\ref{eq:constr_unit_lb}). 
	To solve it, we relax $\|\bm{l}_b\| = 1$ in (\ref{eq:constr_unit_lb}) as $\|\bm{l}_b\|^2 \leq 1$.
	While for (\ref{eq:constr_qb_dist_lb}), by denoting $\bm{l}_b^{t-1}$ as the unit-length feasible solution to (P1-t2) after iteration $t-1$, we deal with it by applying the successive convex approximation (SCA) method via linearizing (\ref{eq:constr_qb_dist_lb}). Specifically, (\ref{eq:constr_qb_dist_lb}) is equivalent to
	\begin{align}\label{eq:quadratic_form_constr}
		\|\bm{l}_b\|^2 - 2 \bm{l}_j^T \bm{l}_b + \|\bm{l}_j\|^2 \geq d_{\text{min}}^2/R^2, \enspace \forall j \in \mathcal{B} \setminus b,
	\end{align}
	where the first term $\|\bm{l}_b\|^2$ can be approximated via its first-order Taylor's approximation at $\bm{l}_b^{t-1}$ as
	\begin{align}\label{eq:Taylor_exp}
		\|\bm{l}_b\|^2 \approx \|\bm{l}_b^{t-1}\|^2 + 2 (\bm{l}_b^{t-1})^T(\bm{l}_b - \bm{l}_b^{t-1}).
	\end{align}
	With the above relaxations, (P1-t2) is 
	transformed into
	\begin{subequations}
		\begin{align}
			(\text{P1-t2-r}): & \max_{\bm{l}_b} \hat{C}(\bm{l}_b) \\
			\text{s.t. } 
			\label{eq:constr_qb_dist_lb_half}
			& 2 (\bm{l}_b^{t-1} - \bm{l}_j)^T \bm{l}_b \geq d_{\text{min}}^2/R^2, \enspace \forall j \in \mathcal{B} \setminus b,\\
			\label{eq:constr_unit_lb_half}
			& \|\bm{l}_b\|^2 \leq 1.
		\end{align}
	\end{subequations}
	Given $\bm{l}_b^{t-1}$ at iteration $t-1$, we then apply the Frank–Wolfe (conditional gradient descent) algorithm 
	to obtain $\bm{l}_b^{t}$. Specifically, we first find $\tilde{\bm{l}}_b^{t-1}$ to maximize the first-order Taylor approximation of $\hat{C}(\bm{l}_b^{t-1})$ within the feasible region by solving
	\begin{subequations}
		\begin{align}
			(\text{P1-t2-c}): & \max_{\bm{l}_b} \nabla \hat{C}(\bm{l}_b^{t-1})^T \bm{l}_b \\
			\text{s.t. } & \text{(\ref{eq:constr_qb_dist_lb_half}) and (\ref{eq:constr_unit_lb_half})},
		\end{align}
	\end{subequations}
	which is a convex optimization problem that can be efficiently solved \cite{boyd2004convex}. The solution to (P1-t2-c) is denoted as $\tilde{\bm{l}}_b^{t-1}$.
	We then decide the step size $\tau^{t-1}$ to get $\bm{l}_b^{t}$ via backtracking, i.e., 
	\begin{align}
		\bm{l}_b^{t} = \bm{l}_b^{t-1} + \tau^{t-1} \left(\tilde{\bm{l}}_b^{t-1} - \bm{l}_b^{t-1}\right), \enspace 0 \leq \tau^{t-1} \leq 1,
	\end{align}
	where the step size $\tau^{t-1}$ is set such that $\bm{l}_b^{t}$ satisfies the Armijo condition \cite{boyd2004convex}.
	Finally, we normalize $\bm{l}_b^{t}$ as the unit-length feasible solution to (P1-t2) after iteration $t$.
	\begin{algorithm}[t]
		\caption{Algorithm for Solving Problem (P1-t)}
		\label{alg:cond_grad_l}
		\begin{itemize}
			\item \textbf{Input}: $B$, $N$, $\lambda$, $T_{\text{ou,l}}$, $\ell$, $\epsilon_{\text{th}}$, $\tau_{\text{init}}$, $\delta$, and $T_{\text{in,l}}$ 	
			
			\begin{algorithmic}[1]
				\STATE Initialize feasible $\{\bm{l}_i\}_{i \in \mathcal{B}}$, $m=0$, and $\bm{u}_b = \left[\frac{\pi}{2},0\right]^T, \forall b \in \mathcal{B}$.
				\WHILE{$m < T_{\text{ou,l}}$}
				\FOR{$b=1$ to $B$}
				\STATE Fix $\{\bm{l}_i\}_{i \in \mathcal{B} \setminus b}$, and set $t \gets 1$, $\bm{l}_b^0 \gets \bm{l}_b$ 
				\WHILE{$t < T_{\text{in,l}}$}
				\STATE Compute the gradient $\nabla \hat{C}(\bm{l}_b^{t-1})^T$ and set $\tau^{t-1} = \tau_{\text{init}}$
				\STATE Obtain $\tilde{\bm{l}}_b^{(t-1)}$ by solving problem (P1-t2-c)
				\STATE Compute  $\bm{l}_b^{t} = \bm{l}_b^{t-1} + \tau^{t-1} \left(\tilde{\bm{l}}_b^{t-1} - \bm{l}_b^{t-1}\right)$
				\WHILE{$\hat{C}(\bm{l}_b^{t}) - \hat{C}(\bm{l}_b^{t-1}) < \ell \tau^{t-1}  \nabla \hat{C}(\bm{l}_b^{t-1})^T (\tilde{\bm{l}}_b^{(t-1)} - \bm{l}_b^{t-1})$}
				\STATE $\tau^{t-1} \gets \delta \tau^{t-1}$
				\STATE Compute $\bm{l}_b^{t} = \bm{l}_b^{t-1} + \tau^{t-1} \left(\tilde{\bm{l}}_b^{t-1} - \bm{l}_b^{t-1}\right)$
				\ENDWHILE
				\STATE $\bm{l}_b^{t} \gets \bm{l}_b^{t}/\|\bm{l}_b^{t}\|$
				\IF{$|\hat{C}(\bm{l}_b^{t}) - \hat{C}(\bm{l}_b^{t-1})| \leq \epsilon_{\text{th}}$}
				\STATE $t \gets t + 1$, \textbf{break}
				\ENDIF
				\STATE $t \gets t + 1$
				\ENDWHILE
				\STATE Update $ \bm{l}_b$ as $\bm{l}_b^{t-1}$
				\ENDFOR
				\STATE $m \gets m + 1$
				\ENDWHILE
			\end{algorithmic}
			\item \textbf{Output}: $\{\bm{l}_i^\star\}_{i \in \mathcal{B}}$ or $\{\bm{t}_i^\star\}_{i \in \mathcal{B}}$ equivalently
		\end{itemize}
	\end{algorithm}
	The above optimization of $\bm{l}_b$ yields $\bm{l}_b^0,\bm{l}_b^1,...$ and terminates when either the convergence criterion of $|\hat{C}(\bm{l}_b^{t}) - \hat{C}(\bm{l}_b^{t-1})| \le \epsilon_{\text{th}}$ is met with $\epsilon_{\text{th}}$ denoting a small positive constant or a predefined maximum inner iteration number $T_{\text{in,l}}$ is reached.

	In the outer iteration, the alternating optimization over $\{\bm{l}_j\}_{j \in \mathcal{B}}$ is repeated until the objective value of (P1-t) converges\footnote{This is ensured since the objective value of (P1-t) is non-decreasing over the iterations due to the gradient-based alternating optimization approach and it is upper-bounded by a finite value.} or a predefined maximum outer iteration number $T_{\text{ou,l}}$ is reached.
	Note that the initial $\{\bm{l}_j\}_{j \in \mathcal{B}}$ in the first round can be set as any feasible solution to (P1-t).
	The overall algorithm to solve (P1-t) is given in Algorithm~\ref{alg:cond_grad_l} and its resulting solution is denoted by $\bm{t}^\star$.

	\subsection{Optimization of $\bm{u}$ with Given $\bm{t}^\star$}\label{sec:u_P1} 
	After we solve (P1-t) and obtain the optimized array positions $\bm{t}^\star$, we proceed to optimize the array local rotations/orientations $\bm{u}$ in (P1) with their  positions fixed as $\bm{t}^\star$, i.e.,
	\begin{subequations}
		\begin{align}
			(\text{P1-u})&:  \max_{\bm{u}} \hat{C}(\bm{t}^\star,\bm{u}) \\
			\label{eq:constr_ub_reflection_overall}
			\text{s.t. } & (\bm{R}(\bm{u}_i)\left[:,3\right])^T  \bm{R}(\bm{t}_i^\star)^T (\bm{l}_j(\bm{t}_j^\star) - \bm{l}_i(\bm{t}_i^\star)) \leq 0, \\
			\nonumber
			& \qquad \qquad \enspace \forall i,j \in \mathcal{B}, j \neq i.
		\end{align}
	\end{subequations}
	Similar to (P1-t), (P1-u) is also solved via alternating optimization, i.e., by optimizing each $\bm{u}_b$ with given ${\{\bm{u}_j\}}_{j \in \mathcal{B} \setminus b}$ iteratively. In each iteration, the optimization of $\bm{u}_b$ under fixed
	${\{\bm{u}_j\}}_{j \in \mathcal{B} \setminus b}$
	is given by 
	\begin{subequations}
		\begin{align}
			(\text{P1-u1})&:  \max_{\bm{u}_b} \hat{C}(\bm{t}^\star,\bm{u}_b) \\
			\label{eq:constr_ub_reflection}
			\text{s.t. }  (&\bm{R}(\bm{u}_b)\left[:,3\right])^T  \bm{R}(\bm{t}_b^\star)^T (\bm{l}_j(\bm{t}_j^\star) - \bm{l}_b(\bm{t}_b^\star)) \leq 0, \forall j \in \mathcal{B} \setminus {b}.
		\end{align}
	\end{subequations}
	Although the constraints in (\ref{eq:constr_ub_reflection}) are non-convex, we can approximately linearize them into convex ones. 
	Denote $\bm{u}_b^{t-1} = \left[\vartheta_b^{t-1},\varphi_b^{t-1}\right]$ as the rotation variable after iteration $t-1$, and $\Delta \bm{u}_b =  \bm{u}_b - \bm{u}_b^{t-1} = \left[\Delta\vartheta_b, \Delta\varphi_b\right]^T$.
	In the $t$-th iteration, we aim to obtain the increment $\Delta \bm{u}_b$ that maximizes the objective function of (P1-u1), which is equivalently expressed as
	\begin{align}
		(\text{P1-u2})&:  \max_{\Delta \bm{u}_b} \hat{C}(\bm{t}^\star,\bm{u}_b^{t-1}\text{+}\Delta \bm{u}_b) \\
		\nonumber
		\text{s.t. } & (\bm{l}_j(\bm{t}_j^\star) -  \bm{l}_b(\bm{t}_b^\star))^T \bm{R}(\bm{t}_b^\star) \bm{R}(\bm{u}_b^{t-1}\text{+}\Delta \bm{u}_b)\left[:,3\right]   \leq 0, \\
		\label{eq:constr_ubT_reflection_e}
		& \qquad \qquad \forall j \in \mathcal{B} \setminus b.
	\end{align}
	Notice that we have 
	\begin{align}
		\nonumber
		\bm{R}(\bm{u}_b^{t-1}\text{+}\Delta \bm{u}_b)\left[:,3\right] =
		\left[
		\arraycolsep=2.5pt\def\arraystretch{1.2}
		\begin{array}{c}
			\cos(\vartheta_b^{t-1}\text{+}\Delta \vartheta_b) \cos(\varphi_b^{t-1}\text{+}\Delta \varphi_b)  \\[1pt]
			\cos(\vartheta_b^{t-1}\text{+}\Delta \vartheta_b) \sin(\varphi_b^{t-1}\text{+}\Delta \varphi_b)  \\[1pt]
			\sin(\vartheta_b^{t-1}\text{+}\Delta \vartheta_b)  \\
		\end{array}\right].
	\end{align}
	When $x$ is small, we have $\cos x \approx 1$ and $\sin x \approx x$. Accordingly, (P1-u2) can be transformed into
	\begin{align}
		(\text{P1-u2-r})&:  \max_{\Delta \bm{u}_b} \hat{C}(\bm{t}^\star,\bm{u}_b^{t-1}\text{+}\Delta \bm{u}_b) \\
		\nonumber
		\text{s.t. }  & (\bm{l}_j(\bm{t}_j^\star) - \bm{l}_b(\bm{t}_b^\star))^T \bm{R}(\bm{t}_b^\star) \left(\bm{A}^{t-1} \Delta \bm{u}_b \text{+} \bm{b}^{t-1}\right)  \leq 0, \\
		\label{eq:constr_ubT_reflection}
		& \qquad \qquad \forall j \in \mathcal{B} \setminus b,
	\end{align}
	where $\bm{A}^{t-1}$ and $\bm{b}^{t-1}$ are based on $\bm{u}_b^{t-1}$, which are given by
	\begin{align}
		\nonumber
		\bm{A}^{t-1} = \left[
		\arraycolsep=1.5pt\def\arraystretch{1.2}
		\begin{array}{cc}
			-\sin \vartheta_b^{t-1} \cos \varphi_b^{t-1}    & \cos \vartheta_b^{t-1} \sin \varphi_b^{t-1}   \\[1pt]
			-\sin \vartheta_b^{t-1} \sin \varphi_b^{t-1}    & \cos \vartheta_b^{t-1} \cos \varphi_b^{t-1}   \\[1pt]
			\cos \vartheta_b^{t-1}    &  0    \\
		\end{array}\right]
	\end{align}
	and $\bm{b}^{t-1} = \left[\cos \vartheta_b^{t-1} \cos \varphi_b^{t-1},\cos \vartheta_b^{t-1} \sin \varphi_b^{t-1},\sin \vartheta_b^{t-1}\right]^T$.
	The constraints in (P1-u2-r) are now convex and (P1-u2-r) can be solved by the conditional gradient descent algorithm, i.e.,
	\begin{align}
		(\text{P1-u2-c})&:  \max_{\Delta \bm{u}_b} \nabla \hat{C}(\bm{t}^\star,\bm{u}_b^{t-1})^T (\bm{u}_b^{t-1}+\Delta \bm{u}_b) \\
		\nonumber
		\text{s.t. }  & (\text{\ref{eq:constr_ubT_reflection}}),
	\end{align}
	which is a convex optimization problem.
	\begin{algorithm}[t]
		\caption{Algorithm for Solving Problem (P1-u)}
		\label{alg:cond_grad_u}
		\begin{itemize}
			\item \textbf{Input}: $B$, $N$, $\lambda$, $T_{\text{ou,u}}$, $\ell$, $\epsilon_{\text{th}}$, $\tau_{\text{init}}$, $\delta$, and $T_{\text{in,u}}$ 	
			
			\begin{algorithmic}[1]
				\STATE With fixed $\{\bm{t}_i^\star\}_{i \in \mathcal{B}}$, set $m=0$, and initial $\bm{u}_b = \left[\frac{\pi}{2},0\right]^T, \forall b \in \mathcal{B}$.
				\WHILE{$m < T_{\text{ou,u}}$}
				\FOR{$b=1$ to $B$}
				\STATE Fix $\{\bm{u}_i\}_{i \in \mathcal{B} \setminus b}$, and set $t \gets 1$, $\bm{u}_b^0 \gets \bm{u}_b$ 
				\WHILE{$t < T_{\text{in,u}}$}
				\STATE Compute the gradient $\nabla \hat{C}(\bm{t}^\star,\bm{u}_b^{t-1})^T$ and set $\tau^{t-1} = \tau_{\text{init}}$
				\STATE Obtain $\Delta \bm{u}_b$ by solving problem (P1-u2-c)
				\STATE Compute  $\bm{u}_b^{t} = \bm{u}_b^{t-1} + \tau^{t-1} \Delta \bm{u}_b$
				\WHILE{$\hat{C}(\bm{t}^\star,\bm{u}_b^{t}) - \hat{C}(\bm{t}^\star,\bm{u}_b^{t-1}) < \ell \tau^{t-1} $  $\nabla \hat{C}(\bm{t}^\star, \bm{u}_b^{t-1})^T \Delta \bm{u}_b$ or $\bm{u}_b^t$ is not feasible}
				\STATE $\tau^{t-1} \gets \delta \tau^{t-1}$
				\STATE Compute  $\bm{u}_b^{t} = \bm{u}_b^{t-1} + \tau^{t-1} \Delta \bm{u}_b$
				\ENDWHILE
				\IF{$|\hat{C}(\bm{t}^\star,\bm{u}_b^{t}) - \hat{C}(\bm{t}^\star,\bm{u}_b^{t-1})| \leq \epsilon_{\text{th}}$}
				\STATE $t \gets t + 1$, \textbf{break}
				\ENDIF
				\STATE $t \gets t + 1$
				\ENDWHILE
				\STATE Update $ \bm{u}_b$ as $\bm{u}_b^{t-1}$
				\ENDFOR
				\STATE $m \gets m + 1$
				\ENDWHILE
			\end{algorithmic}
			\item \textbf{Output}: $\{\bm{u}_i^\star\}_{i \in \mathcal{B}}$
		\end{itemize}
	\end{algorithm}
	With $\Delta \bm{u}_b$ at hand, the exact step size $\tau^{t-1} \in \left[0,1\right]$ along the direction $\Delta \bm{u}_b$ can be determined via backtracking, such that $\bm{u}_b^{t} = \bm{u}_b^{t-1} + \tau^{t-1} \Delta \bm{u}_b$ satisfies the Armijo condition and (\ref{eq:constr_ub_reflection}). This is always feasible when $\tau^{t-1} \to 0$. 
	%	The update $\bm{u}_b^{t}$ is obtained after iteration $t$.
	The above optimization of $\bm{u}_b$ yields $\bm{u}_b^0, \bm{u}_b^1, ...$ and terminates when either the convergence criterion of $|\hat{C}(\bm{u}_b^{t}) - \hat{C}(\bm{u}_b^{t-1})| \le \epsilon_{\text{th}}$ is met or the maximum inner iteration number $T_{\text{in,u}}$ is reached.

	In the outer iteration, the alternating optimization over $\{\bm{u}_j\}$ is repeated until the objective value of (P1-u) converges or a predefined maximum outer iteration number $T_{\text{ou,u}}$ is reached.
	Note that we set the initial $\bm{u}_j = \left[\pi/2,0\right]^T, \forall j \in \mathcal{B}$ in the first round, i.e., without any local rotation.
	The overall algorithm to solve (P1-u) is presented in Algorithm~\ref{alg:cond_grad_u}.
	
	As a summary, the overall algorithm to solve (P1) first executes Algorithm~\ref{alg:cond_grad_l} to solve (P1-t) and obtain $\bm{t}^\star$, then executes Algorithm~\ref{alg:cond_grad_u} to solve (P1-u) based on the obtained $\bm{t}^\star$ and attain $\bm{u}^\star$.

	\section{Array Position and Rotation Optimization for Airway Sensing}\label{sec:sensing_problem_formulation}
	
	For airway sensing, we aim to maximize the minimum received sensing signal power along the airway segment via jointly optimizing $\bm{t}$, $\bm{u}$, and transmit covariance matrix $\bm{R}_d$ under the constraints (\ref{eq:dist_constr}) and (\ref{eq:constr_signal_reflection}) as well as the sum-power constraint (\ref{eq:sum-power}). The optimization problem is formulated as\footnote{For simplicity, we present the problem here considering one single airway, while the formulated problem and proposed algorithm for solving it in this section can be extended to the general case with multiple airways by maximizing the minimum received power over them.}
	\begin{subequations}
		\begin{align}
			(\text{P}2): & \max\limits_{\bm{t}, \bm{u}, \bm{R}_d} \enspace \min_{\xi \in \left[0,1\right]}  \bm{h}_\xi^T(\bm{t},\bm{u}) \bm{R}_d \bm{h}_\xi^*(\bm{t},\bm{u}) \\
			\text { s.t. } 
			\label{equ:sum-max-min}
			& (\bm{R}(\bm{u}_i)\left[:,3\right])^T  \bm{R}(\bm{t}_i)^T (\bm{l}_j(\bm{t}_j) - \bm{l}_i(\bm{t}_i)) \leq 0, \\
			& \|\bm{l}_i(\bm{t}_i) - \bm{l}_j(\bm{t}_j)\| \geq d_{\text{min}}/R, \enspace \forall i,j \in \mathcal{B}, j \neq i,\\
			& \operatorname{tr}(\bm{R}_d) \leq P_0.
		\end{align}
	\end{subequations}
	%	Notice that (P2) is highly non-convex.
	To tackle this non-convex optimization problem, we first fix $\bm{R}_d$ and optimize the positions $\bm{t}$ and rotations $\bm{u}$ of all the HT-6DMA arrays, followed by the optimization of $\bm{R}_d$ based on the optimized $\bm{t}$ and $\bm{u}$. The procedure of optimizing  $\bm{t}$ and $\bm{u}$ in (P2) is similar as that given in Section \ref{sec:Uplink_com}, while their difference is due to the fact that the objective function of (P2) is non-differentiable and thus more difficult to handle.
	%	\begin{align}\label{eq:obj_P2_sensing}
		%		\min_{\xi \in \left[0,1\right]} \mathcal{P}(\theta(\xi),\phi(\xi)) = \min_{\xi \in \left[0,1\right]} \bm{h}_\xi^T(\bm{t},\bm{u}) \bm{R}_d \bm{h}_\xi^*(\bm{t},\bm{u})
		%	\end{align}
	In the following, we first optimize $\bm{t}$ and $\bm{u}$ with $\bm{R}_d$ fixed by transforming the non-differentiable objective function of (P2) into a differentiable one. Then, we proceed to optimize $\bm{R}_d$ based on the optimized $\bm{t}$ and $\bm{u}$.

	\subsection{Optimization of $\bm{t}$ and $\bm{u}$ with Fixed $\bm{R}_d$}\label{eq:Sensing_opt_tu}
	We set $\bm{R}_d$ as $\bm{R}_d = \frac{P_0}{N B} \bm{I}_{NB}$, which satisfies the sum-power constraint in (\ref{eq:sum-power}) and allows us to focus on optimizing $\bm{t}$ and $\bm{u}$.
	Accordingly, (P2) with $\bm{R}_d =  \frac{P_0}{N B} \bm{I}_{NB}$ is given by
	\begin{subequations}
		\begin{align}
			\label{eq:obj_P21}
			(\text{P2-1}): & \max\limits_{\bm{t}, \bm{u}} \enspace \min_{\xi \in \left[0,1\right]}  \frac{P_0}{N B} \|\bm{h}_\xi(\bm{t},\bm{u})\|^2 \\
			\text { s.t. } 
			\label{equ:rot_sensing}
			& (\bm{R}(\bm{u}_i)\left[:,3\right])^T  \bm{R}(\bm{t}_i)^T (\bm{l}_j(\bm{t}_j) - \bm{l}_i(\bm{t}_i)) \leq 0, \\
			\label{eq:pos_sensing}
			& \|\bm{l}_i(\bm{t}_i) - \bm{l}_j(\bm{t}_j)\| \geq d_{\text{min}}/R, \enspace \forall i,j \in \mathcal{B}, j \neq i.
		\end{align}
	\end{subequations}
	The objective function in (\ref{eq:obj_P21}) is not differentiable. To tackle this issue, we first discretize the continuous interval $\left[0,1\right]$ into a discrete set  containing $D$ discrete points equally spaced between 0 and 1, which is denoted as 
	\begin{align}\label{eq:Set_xi}
		\bm{\Xi} \triangleq \left\{0,\frac{1}{D},...,\frac{D-1}{D}\right\}.
	\end{align}
	When $D$ is large enough, we have
	\begin{align}
		\min_{\xi \in \left[0,1\right]}\frac{P_0}{N B} \|\bm{h}_\xi(\bm{t},\bm{u})\|^2 & = \min_{\xi \in \left[0,1\right]} f(\bm{t},\bm{u},\xi)\\ 
		& \approx \min_{\xi \in \bm{\Xi}} f(\bm{t},\bm{u},\xi) \triangleq f(\bm{t},\bm{u}).
	\end{align}
	Then, we apply the entropic regularization method \cite{polyak1988smooth} to further approximate $f(\bm{t},\bm{u})$ via $f_a(\bm{t},\bm{u})$,  which is defined as
	%	\begin{align}
		%		f(\bm{t},\bm{u}) \triangleq \min_{\xi \in \left[0,1\right]}\frac{P_0}{N B} \|\bm{h}_\xi(\bm{t},\bm{u})\|^2 = \min_{\xi \in \left[0,1\right]} f(\bm{t},\bm{u},\xi)
		%	\end{align}
	%	as 
	\begin{align}
		f_a(\bm{t},\bm{u}) \triangleq -\frac{1}{\beta} \ln \left(\sum\nolimits_{\xi \in \bm{\Xi}} e^{-\beta f(\bm{t},\bm{u},\xi)}\right),
	\end{align}
	where $\beta$ is a hyper-parameter used to tune the accuracy of the approximation.  It is easy to show that $f_a(\bm{t},\bm{u}) \leq f(\bm{t},\bm{u})$ and when $\beta \to \infty$, $f_a(\bm{t},\bm{u}) \to f(\bm{t},\bm{u})$. That is, instead of maximizing $f(\bm{t},\bm{u})$ directly, we aim to maximize the surrogate function $f_a(\bm{t},\bm{u})$, which is a lower bound of $f(\bm{t},\bm{u})$, and the gap between them becomes negligible when $\beta$ is sufficiently large. With the above two approximations, the objective function of (P2-1) becomes differentiable and (P2-1) is approximated as
	\begin{subequations}
		\begin{align}
			(\text{P2-2})&:  \max\limits_{\bm{t},\bm{u}} \enspace f_a(\bm{t},\bm{u}) \\
			\text {s.t. } 
			\label{equ:rot_sensing_approx}
			& (\bm{R}(\bm{u}_i)\left[:,3\right])^T  \bm{R}(\bm{t}_i)^T (\bm{l}_j(\bm{t}_j) - \bm{l}_i(\bm{t}_i)) \leq 0, \\
			\label{eq:pos_sensing_approx}
			& \|\bm{l}_i(\bm{t}_i) - \bm{l}_j(\bm{t}_j)\| \geq d_{\text{min}}/R, \enspace \forall i,j \in \mathcal{B}, j \neq i.
		\end{align}
	\end{subequations}
	(P2-2) can then be solved using the same approach discussed in Section \ref{sec:Uplink_com} for solving (P1) with the objective function replaced by $ f_a(\bm{t},\bm{u})$.
	%	 Specifically, same as Algorithm \ref{alg:cond_grad_l} in Section \ref{sec:P1-t_opt_t}, $\bm{t}$ is first optimized in an alternating manner while $\bm{u}$ is fixed with $\bm{u}_i = \left[\pi/2,0\right]^T, \forall i \in \mathcal{B}$. With optimized $\bm{t}^\star$, $\bm{u}$ is optimized also in an alternating fashion, as shown in Algorithm \ref{alg:cond_grad_u} in Section \ref{sec:u_P1}, and the optimized rotations $\bm{u}^\star$ can then be obtained.
	
	\subsection{Optimization of $\bm{R}_d$ with Fixed $\bm{t}^\star$ and $\bm{u}^\star$}
	
	After the positions and rotations of all the arrays are obtained as $\bm{t}^\star$ and $\bm{u}^\star$ by solving (P2-2), we proceed to optimize $\bm{R}_d$ in (P2). The problem is formulated as
	\begin{subequations}
		\begin{align}
			(\text{P2-3})&:  \max\limits_{ \bm{R}_d, \chi} \enspace \chi \\
			\label{equ:max-min_sim_Rd_only}
			\text {s.t. } & \bm{h}_\xi^T(\bm{t}^\star,\bm{u}^\star) \bm{R}_d \bm{h}_\xi^*(\bm{t}^\star,\bm{u}^\star) \geq \chi, \enspace \forall \xi \in \bm{\Xi}, \\
			\label{equ:sum-max-min_sim_Rd_only}
			& \text{tr}(\mv{R}_d) \leq P_0,
		\end{align}
	\end{subequations}
	where $\chi = \min_{\xi \in \bm{\Xi}} \bm{h}_\xi^T(\bm{t}^\star,\bm{u}^\star) \bm{R}_d \bm{h}_\xi^*(\bm{t}^\star,\bm{u}^\star)$ is introduced as an auxiliary variable. Here, we also use the discrete set $\bm{\Xi}$ given in (\ref{eq:Set_xi}) instead of the continuous interval $\left[0,1\right]$ for the sake of optimization.
	Notice that (P2-3) is a standard separable semidefinite programming (SSDP), which is convex and can be solved optimally by numerical solver like CVX \cite{grant2014cvx}.
	
	\section{Numerical Results}\label{sec:numerical_result}
	
\begin{figure}[t]
		\centering
		\setlength{\abovecaptionskip}{+4mm}
		\setlength{\belowcaptionskip}{+1mm}
		\subfigure[Radiation pattern $A_h(\phi)$ in dB.]{ \label{fig:Horizontal_radiation_pattern_Ah}
			\includegraphics[width=1.672in]{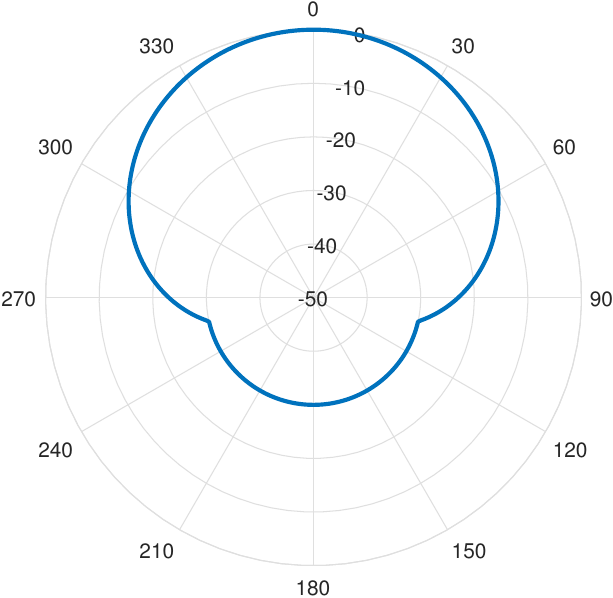}}
		\subfigure[3D directional gain $A(\theta,\phi)$ in dB.]{ \label{fig:Radiation_pattern_3D}
			\includegraphics[width=1.672in]{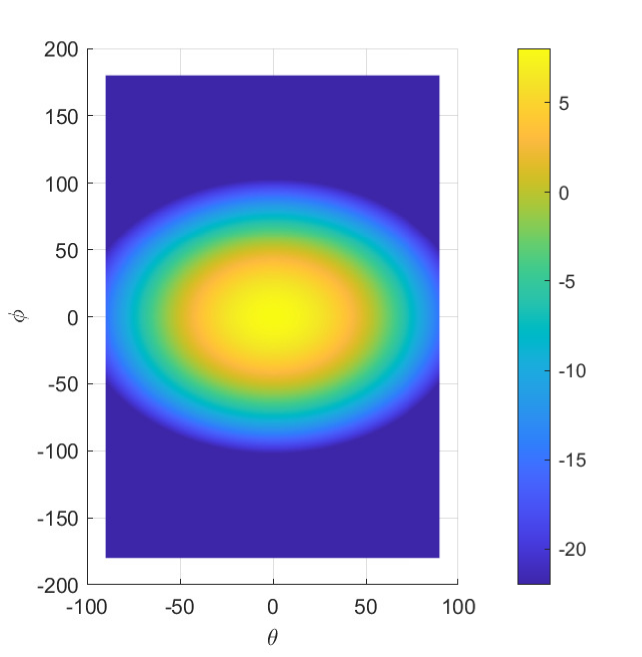}}
		\caption{Radiation pattern specified in 3GPP \cite{3gpp2018study}.}
	\end{figure}

	In this section, we provide numerical results to validate the performance of our proposed algorithms based on the HT-6DMA model, as compared to other benchmark schemes in both uplink communication and airway sensing scenarios.
	Without loss of generality, we consider each BS array as a UPA with half-wavelength antenna spacing and set the radius of the sphere $\mathcal{C}$ as $R=1$ meter (m). The directional antenna gain $A(\theta,\phi)$ in dB in (\ref{eq:directional_gain}) is
	given as \cite{3gpp2018study}
	\begin{align}
		A(\theta,\phi) = G_{\text{max}} - \min \left\{-\left[A_h(\phi)+A_v(\theta)\right],G_s\right\},
	\end{align}
	with $A_h(\phi) = - \min \{12 \left(\frac{\phi}{\phi_{\text{3dB}}}\right)^2, G_s\}$ and $A_v(\theta) = - \min \{12 \left(\frac{\theta}{\theta_{\text{3dB}}}\right)^2, G_v\}$, where
	$\phi_{\rm{3dB}}$ and $\theta_{\rm{3dB}}$ are the 3-dB beamwidth in the horizontal and vertical planes, respectively, which are both set to be 65$^{\circ}$. $G_s$ and $G_v$ denote the front-back ratio and sidelobe level limit, respectively, with	
	$G_s = G_v = 30$ dB. $G_{\rm{max}}$ is the maximum directional gain of each antenna element in the main lobe direction and is set to be $8$ dBi \cite{3gpp2018study}.
	The corresponding horizontal radiation pattern $A_h(\phi)$ and 3D directional gain $A(\theta,\phi)$ are shown in Fig. \ref{fig:Horizontal_radiation_pattern_Ah} and Fig. \ref{fig:Radiation_pattern_3D}, respectively.
	Throughout this section, to compare with the proposed HT-6DMA (with both position adjustment and rotation adjustments), the following benchmarks schemes are considered:
	\begin{itemize}
		\item \textbf{FPA}: The BS has 3 sector antenna arrays equally placed along the equator of $\mathcal{C}$, each with $\lfloor \frac{N B}{3} \rfloor$ antennas. The orientation of each array is fixed with $\bm{u}_b = \left[\frac{5\pi}{12},0\right]^T$.
		\item \textbf{Existing 6DMA \cite{shao20246d}}: $B$, $N$, and $\mathcal{C}$ in this benchmark are set to be the same as those for the HT-6DMA BS.
		The array positions and rotations are jointly optimized by the proposed algorithm in \cite{shao20246d}.
		\item \textbf{HT-6DMA with position adjustment (PA)}: Only the array positions are optimized by our proposed algorithm (i.e., Algorithm \ref{alg:cond_grad_l}) without any local rotation adjustment.
		\item \textbf{HT-6DMA with rotation adjustment (RA)}: The arrays are  uniformly placed on the sphere and only their rotations $\bm{u}$ are optimized by our proposed algorithm (i.e., Algorithm \ref{alg:cond_grad_u}).
	\end{itemize}
	%	While our proposed HT-6DMA algorithm first optimizes array position (i.e., Algorithm \ref{alg:cond_grad_l}), followed by array local rotation optimization (i.e., Algorithm \ref{alg:cond_grad_u}), i.e., HT-6DMA with both PA and RA. In the following, we use HT-6DMA to denote the proposed scheme for simplicity.
	
	\subsection{Uplink Communication Scenario}\label{sec:num_uplink}
	In this subsection, we present the simulation results for the uplink communication scenario.
	Similar to \cite{shao20246d}, we adopt the general non-homogeneous Poisson process (NHPP) to model the user spatial distribution. All users are assumed to be distributed within a 3D coverage region denoted as $\mathcal{A}$, which is a 3D spherical annulus with radial distances ranging
	from 50 m to 120 m from the center of the BS.
	Within $\mathcal{A}$, there exist three hotspot areas, $\mathcal{A}_1$, $\mathcal{A}_2$, and $\mathcal{A}_3$, 
	which are three spheres centered at (30, $-$60, $-$50) m, ($-$40, 0, 60) m, and (0, 100, 20) m, respectively, with the same radii of 15 m. The region excluding these hotspots is denoted as $\mathcal{A}_0$.
	The number of users within each region in each channel realization is denoted by $K_i, i \in \{0,1,2,3\}$, which are all modeled as Possion random variables, with their probability mass functions (PMFs) given by
	\begin{equation} 
		\Pr[K_i = \tilde{K}] = \frac{\mu_i^{\tilde{K}} e^{-\mu_i}}{\tilde{K}!}, \tilde{K} = 0, 1, \dots, \forall i \in \{0,1,2,3\},
	\end{equation} 
	where $\mu_i$ represents the mean of $K_i$. We assume that $\mu_1 = \mu_2 = \mu_3$ and the users are uniformly distributed within each region. To quantify the homogeneous level of the user distribution, we define the homogeneous ratio $\eta$ as $\eta = \mu_0/\left(\sum_{i=0}^{3} \mu_i\right).$
	\begin{figure*}[t]
		\centering
		\setlength{\abovecaptionskip}{+2mm}
		\setlength{\belowcaptionskip}{+1mm}
		\subfigure[User spatial distribution ($\eta = 0$).]{ \label{fig:User_dist_2}
			\includegraphics[width=1.7in]{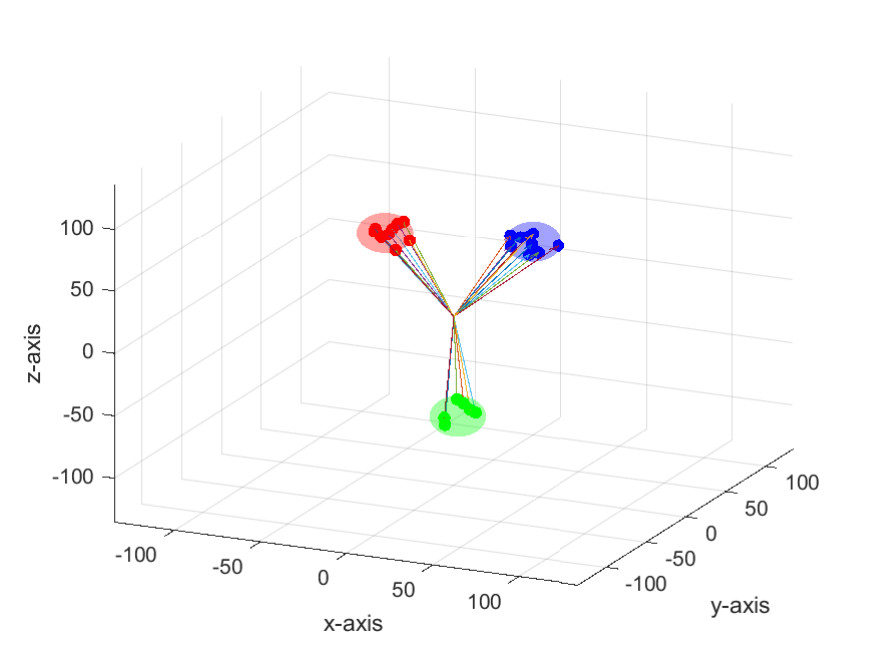}}
		\subfigure[FPA setup with $B=3, N = 21$.]{ \label{fig:FPA_setup}
			\includegraphics[width=1.7in]{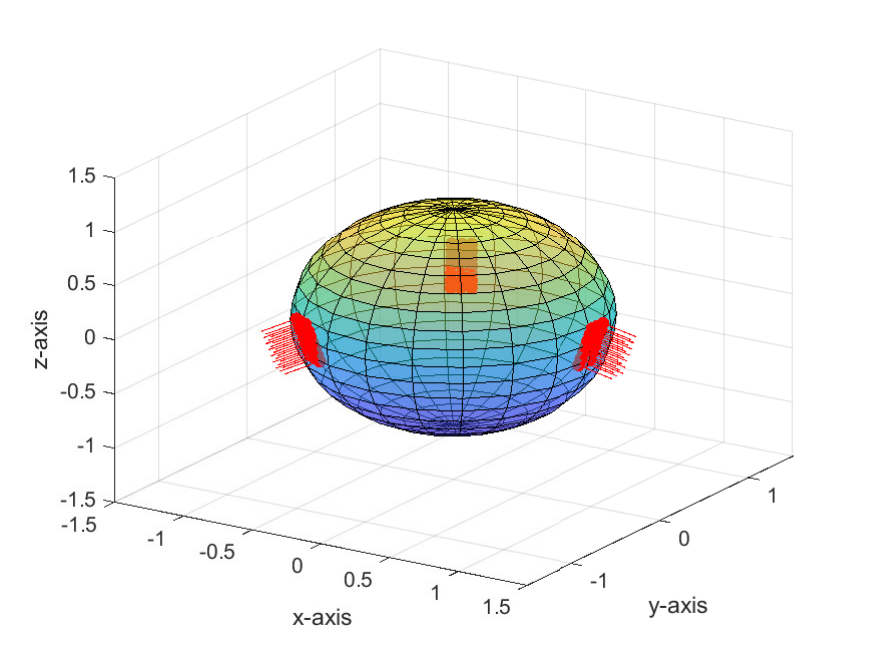}}
		\subfigure[Initial setup for HT-6DMA and existing 6DMA.]{ \label{fig:6DMA_init_2}
			\includegraphics[width=1.7in]{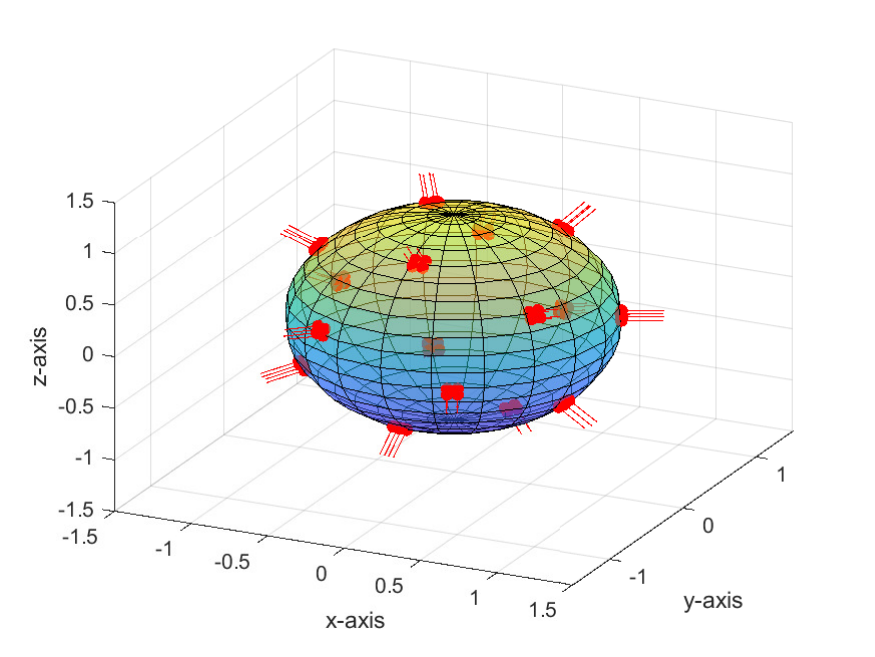}}
		\subfigure[Convergence behavior.]{ \label{fig:6DMA_alg_comp} % Add FPA is left later
			\includegraphics[width=1.7in]{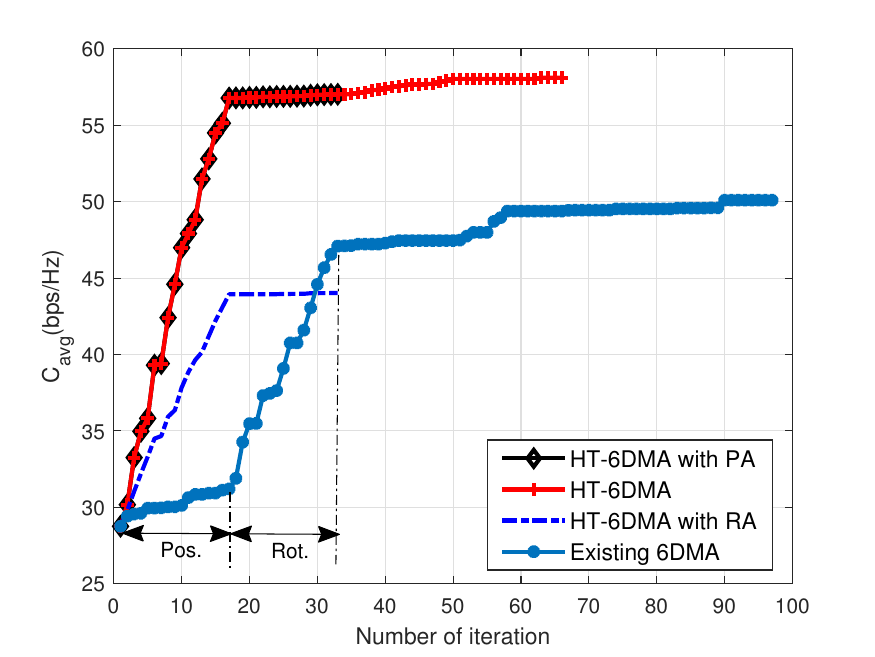}}
		\subfigure[Existing 6DMA \cite{shao20246d}.]{ \label{fig:6DMA_opt_result_pre}
			\includegraphics[width=1.7in]{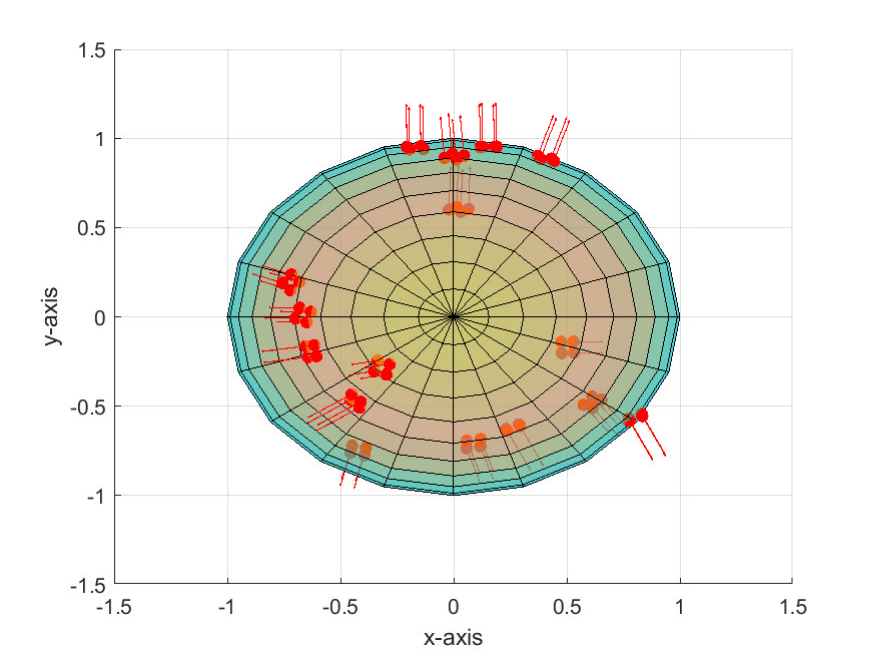}}	
		\subfigure[HT-6DMA with PA.]{ \label{fig:SMA_opt}
			\includegraphics[width=1.7in]{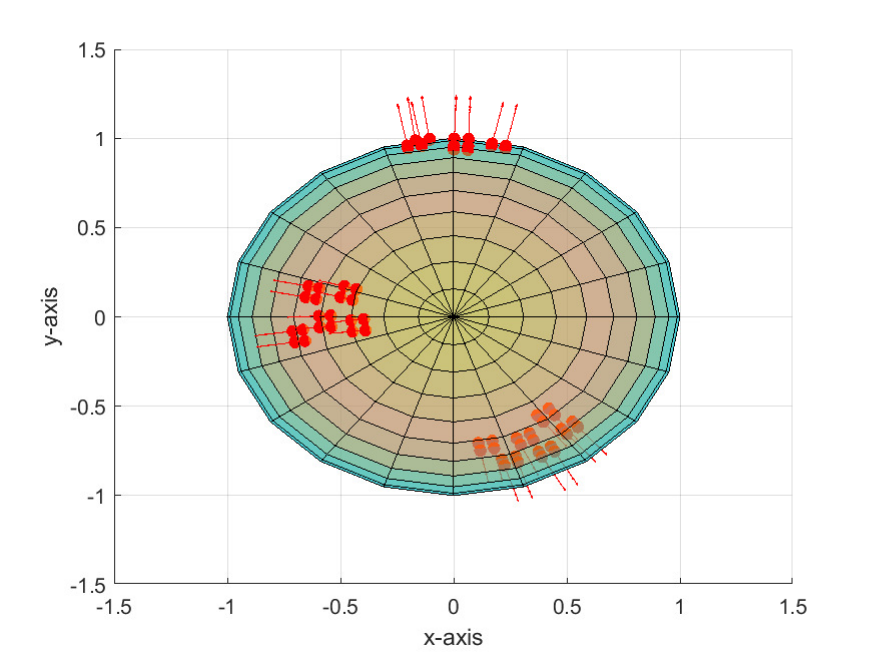}}
		\subfigure[HT-6DMA with RA.]{ \label{fig:SMA_opt_rot}
			\includegraphics[width=1.7in]{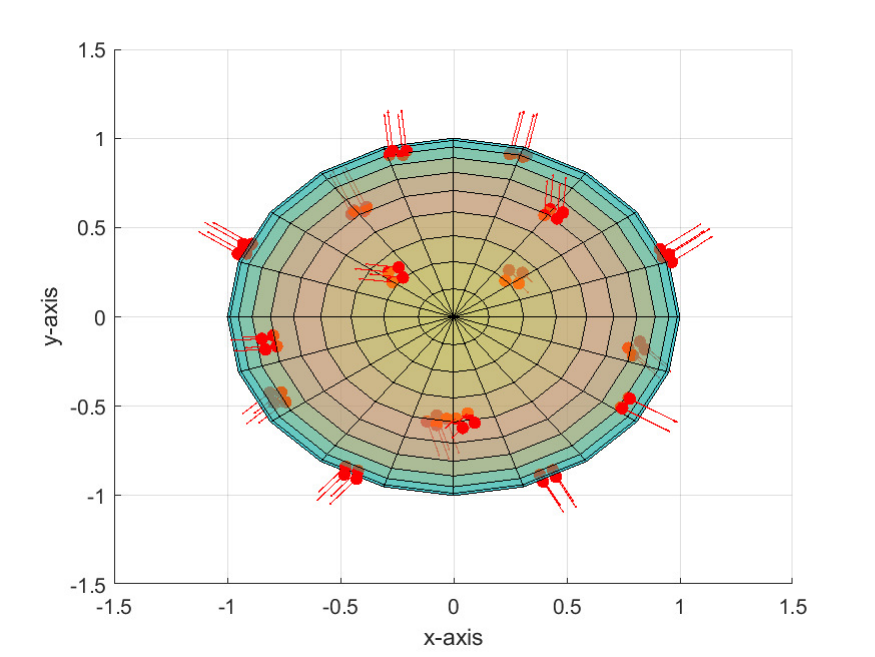}}
		\subfigure[HT-6DMA.]{ \label{fig:SMA_opt_3D}
			\includegraphics[width=1.7in]{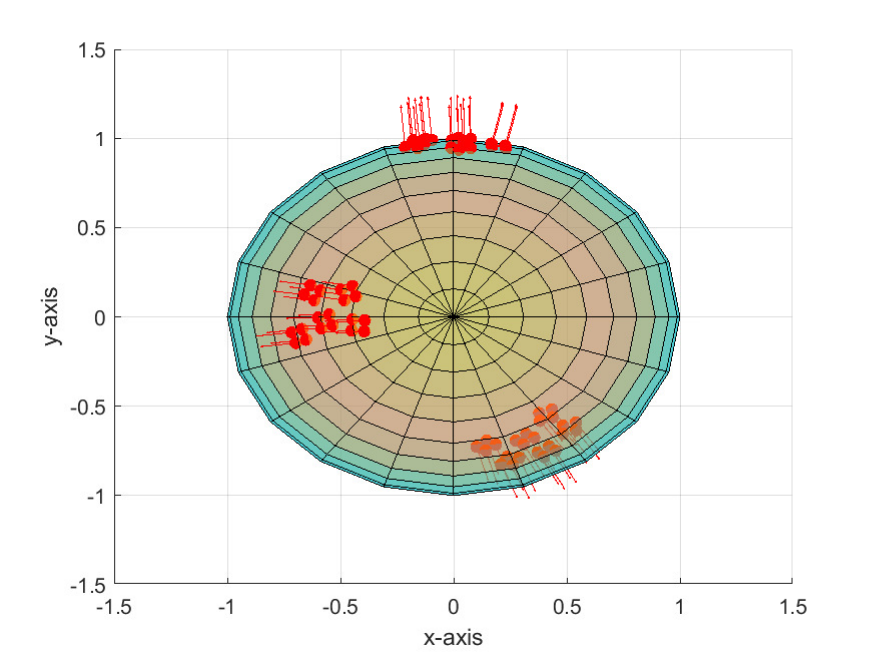}}
		\caption{Simulation and optimization results in uplink communication scenario.}
		\label{fig:uplink_comm_scenario}
	\end{figure*}
	Besides, we perform $S = 100$ MC realizations and set $\mu = \sum_{i=0}^{3} \mu_i = 24$, the average noise power $\sigma^2 = -50$ dBm, the carrier frequency $f_c = 2.4$ GHz, and the transmit power of each user $p = 30$ mW.

	\begin{figure}[t]
		\centering
		\setlength{\abovecaptionskip}{+4mm}
		\setlength{\belowcaptionskip}{+1mm}
		\subfigure[Sum rate comparison.]{ \label{fig:network_cap_comp}
			\includegraphics[width=1.672in]{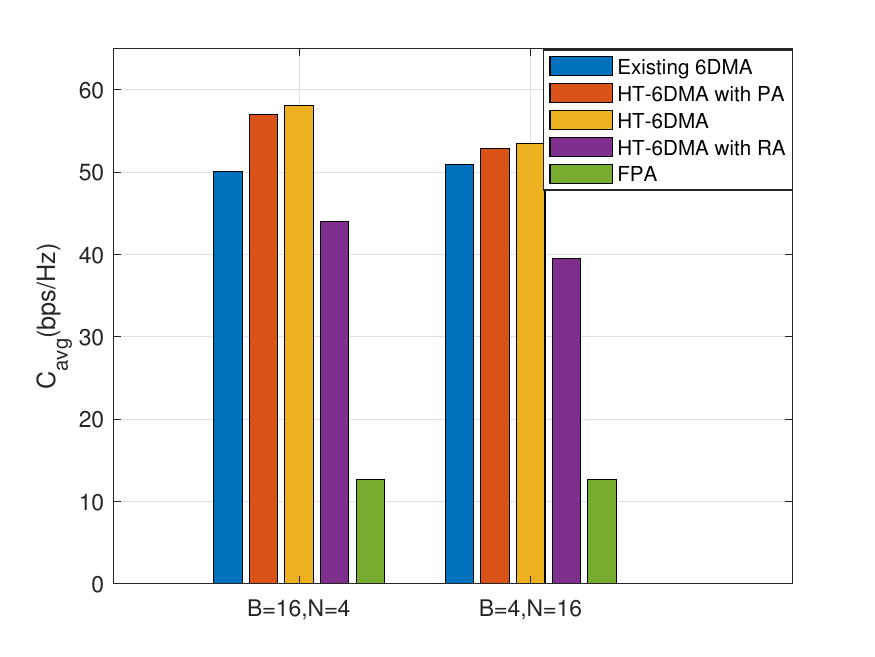}}
		\subfigure[Execution time comparison.]{ \label{fig:Exe_time_comp}
			\includegraphics[width=1.672in]{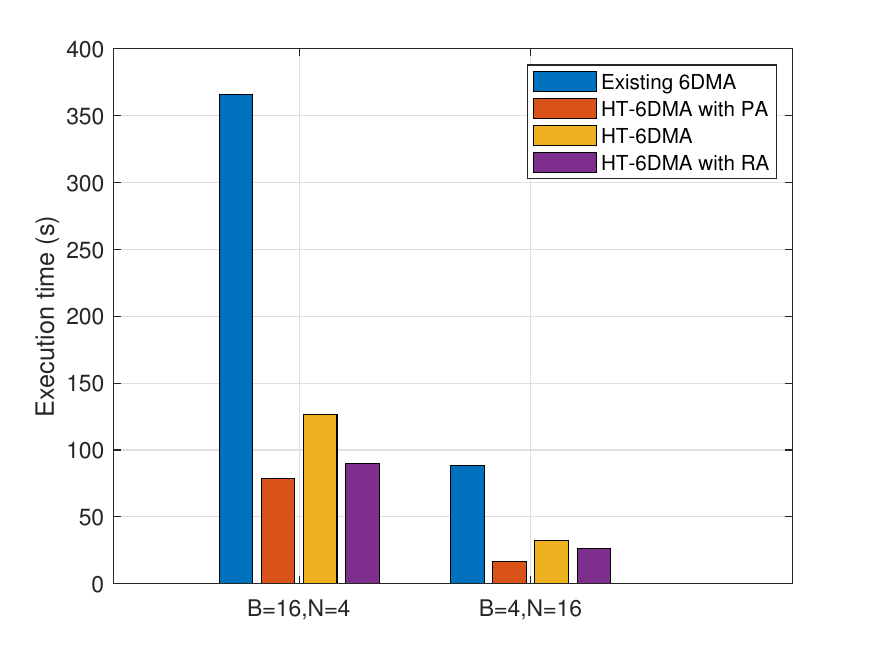}}
		\caption{Performance and complexity comparison in uplink communication scenario.}
		\label{fig:uplink_comm_general_comp}
	\end{figure}
	In Fig. \ref{fig:uplink_comm_scenario}, we compare the performance of the proposed HT-6DMA with the existing 6DMA and FPA under the user spatial distribution shown in Fig. \ref{fig:User_dist_2}, where we set $\epsilon_{\text{th}} = 5 \times 10^{-4}$, $T_{\text{in}} = 50$, and $(T_{\text{ou,l}},T_{\text{ou,u}}) = (2,2)$. 
	For fair comparison, the proposed HT-6DMA and existing 6DMA use the same initial variable values, as shown in Fig. \ref{fig:6DMA_init_2}, while the FPA setup is also shown in Fig. \ref{fig:FPA_setup}. In Fig. \ref{fig:6DMA_alg_comp}, we show the convergence behaviors of the proposed algorithms for HT-6DMA and that for the existing 6DMA with $(B,N) = (16,4)$. Notice that the number of iterations in Fig. \ref{fig:6DMA_alg_comp} corresponds to the number of updates for the position or rotation variable of each array, e.g., the first $B=16$ iterations correspond to the array position updates, followed by the second $B=16$ array orientation updates in the case of existing 6DMA, as highlighted in Fig. \ref{fig:6DMA_alg_comp}.
	It is observed that the proposed HT-6DMA (with both PA and RA) converges to a better local optimum within fewer iterations than the existing 6DMA.
	This is mainly due to two reasons. First, when optimizing the array positions in the existing 6DMA algorithm \cite{shao20246d}, the array orientations w.r.t. the global CCS remain unchanged, while
	in the HT-6DMA algorithm, although there is no array rotation w.r.t. the local SCS when optimizing the array positions, the array orientations w.r.t. the global SCS will change with the array positions moving on the spherical surface, leading to a more efficient search for optimal array positions/rotations. Second, in the case of HT-6DMA, the optimizations of array position and array orientation are decoupled with simpler constraints as compared to those in the existing 6DMA algorithm \cite{shao20246d}, which leads to faster convergence to a better solution. 
	Furthermore, it is observed from Fig. \ref{fig:6DMA_alg_comp} that the performance gain by HT-6DMA mostly comes from the position optimization on the sphere instead of the local rotation optimization since the performance gap of the HT-6DMA with PA from the HT-6DMA (with both PA and RA) is very small. It is also shown in Fig. \ref{fig:6DMA_alg_comp} that although the
	scheme of HT-6DMA with RA can improve the sum rate over the FPA scheme, its performance is still much inferior to the other two HT-6DMA schemes, due to the lack of adjustment of array positions. Finally, the optimized array positions and rotations viewed from above by different schemes are shown in Fig. \ref{fig:6DMA_opt_result_pre}-\ref{fig:SMA_opt_3D} for comparison. It is worth noticing that the differences between the HT-6DMA with PA in Fig. \ref{fig:SMA_opt} and the HT-6DMA (with both PA and RA) in Fig. \ref{fig:SMA_opt_3D} is small, which is in accordance with the small performance gap between these two schemes.
	
	Next, we compare the HT-6DMA with other benchmarks in terms of the achievable sum rate as well as algorithm execution time under two setups with different $B$ and $N$ values in Fig. \ref{fig:network_cap_comp} and \ref{fig:Exe_time_comp}, respectively, where the setup with $B = 16, N = 4$ is already considered in Fig. \ref{fig:uplink_comm_scenario}.
	It is clearly seen from Fig. \ref{fig:network_cap_comp} that the existing 6DMA and the HT-6DMA schemes all achieve much higher sum rates than the FPA scheme due to their adaptability to user spatial distribution.
	It is also observed in Fig. \ref{fig:network_cap_comp} that when keeping the total number of antennas fixed, all the HT-6DMA schemes achieve higher rates when $B$ is larger, since more arrays are tunable, which makes the design more flexible. 
	However, with more variables to design and more constraints to consider when $B=16$, the execution time for all 6DMA schemes becomes longer than that with $B=4$, as shown in Fig. \ref{fig:Exe_time_comp}. 
	In particular, the existing 6DMA algorithm requires much more time in finding the solution when $B=16$, but the resultant sum rate is even slightly worse than that with $B=4$, as shown in Fig. \ref{fig:network_cap_comp}. This is possibly due to its more susceptibility to bad local optimum when $B$ is larger.
	In contrast, it is observed that the HT-6DMA algorithm achieves better performance with less execution time than the existing 6DMA algorithm under both setups. Finally, it is observed from Fig. \ref{fig:network_cap_comp} that even for the new setup with smaller $B$ ($B = 4$), the performance gain of HT-6DMA still mostly comes from the position optimization on the sphere rather than the local rotation optimization.
	%	\begin{figure}[t]
		%		\centering
		%		\setlength{\abovecaptionskip}{+4mm}
		%		\setlength{\belowcaptionskip}{+1mm}
		%		\subfigure[Sum rate and gain versus $d_{\text{min}}$.]{ \label{fig:Gap_d_min}
			%			\includegraphics[width=1.6725in]{./figures/Gap_d_min2.eps}}
		%		\subfigure[Gain versus UPA geometry.]{ \label{fig:Gap_shape}
			%			\includegraphics[width=1.6725in]{./figures/Gap_shape.eps}}
		%		\caption{The sum rate performance gain of array local rotation with HT-6DMA versus $d_{\text{min}}$.}
		%		\label{fig:uplink_comm_gap_comp}
		%	\end{figure}
	\begin{figure}[t]
		\centering
		\includegraphics[width=2.1in]{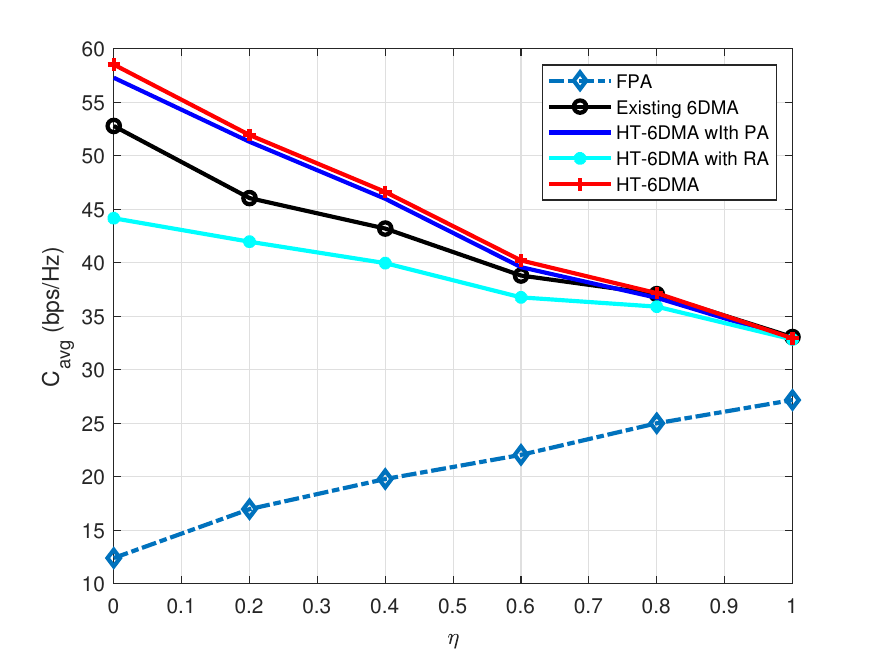}
		\centering
		\caption{Achievable sum rate versus $\eta$.}
		\label{fig:Cap_vs_xi}
	\end{figure}
	\begin{figure}[t]
		\centering
		\includegraphics[width=2.1in]{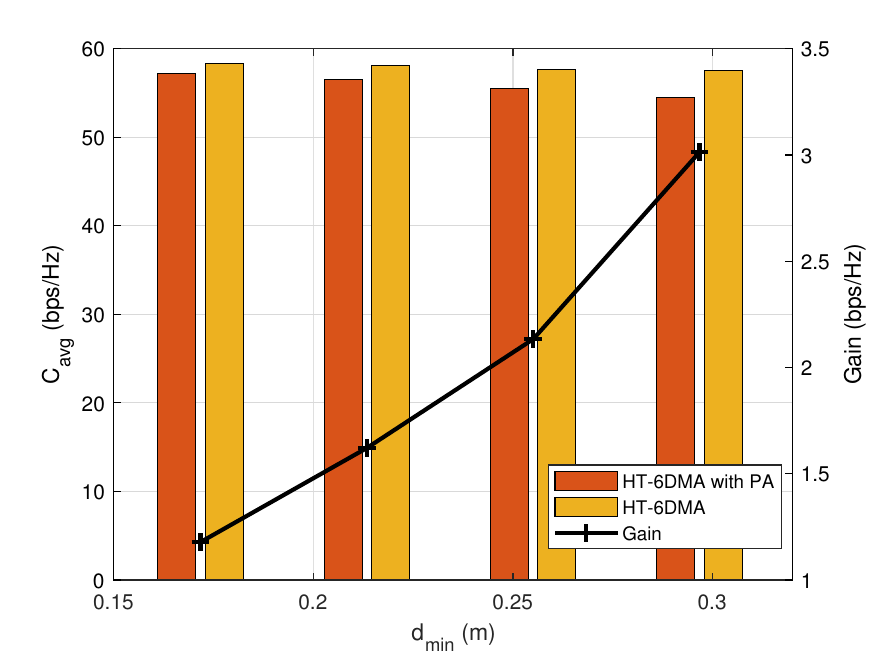}
		\centering
		\caption{The sum rate performance comparison and the gain of array local rotation with HT-6DMA versus $d_{\text{min}}$.}
		\label{fig:Gap_d_min}
	\end{figure}
	
	\begin{figure*}[t]
		\centering
		\setlength{\abovecaptionskip}{+4mm}
		\setlength{\belowcaptionskip}{+1mm}
		\subfigure[Two airways.]{ \label{fig:Airway_3D}
			\includegraphics[width=1.7in]{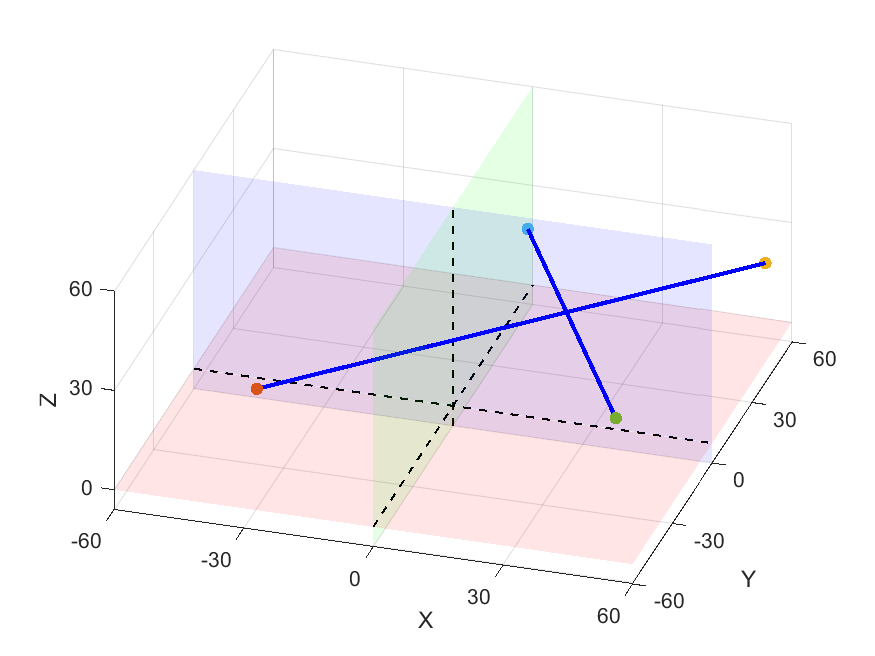}}
		\subfigure[Convergence behavior of $f_a$.]{ \label{fig:Convergence_curve_sensing} 
			\includegraphics[width=1.7in]{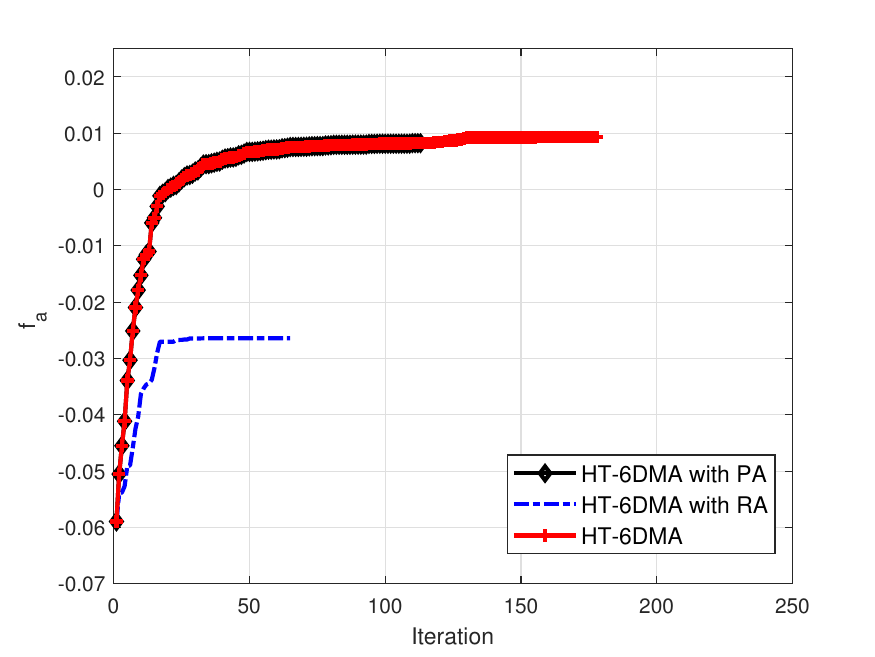}}
		\subfigure[HT-6DMA with RA.]{ \label{fig:Airway_sensing_opt_SMA_1D}
			\includegraphics[width=1.7in]{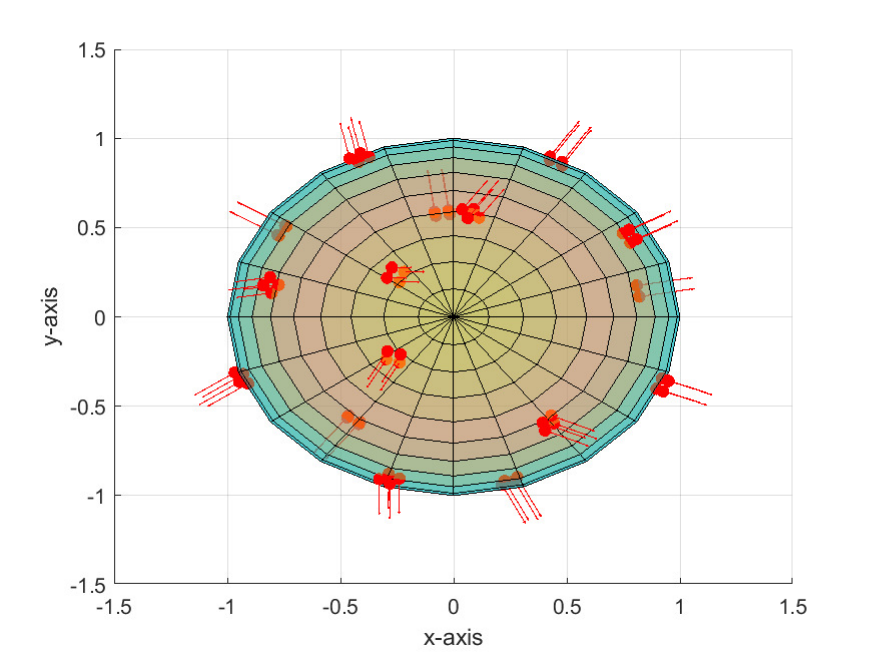}}
		\subfigure[HT-6DMA with PA.]{ \label{fig:Airway_sensing_opt_SMA}
			\includegraphics[width=1.7in]{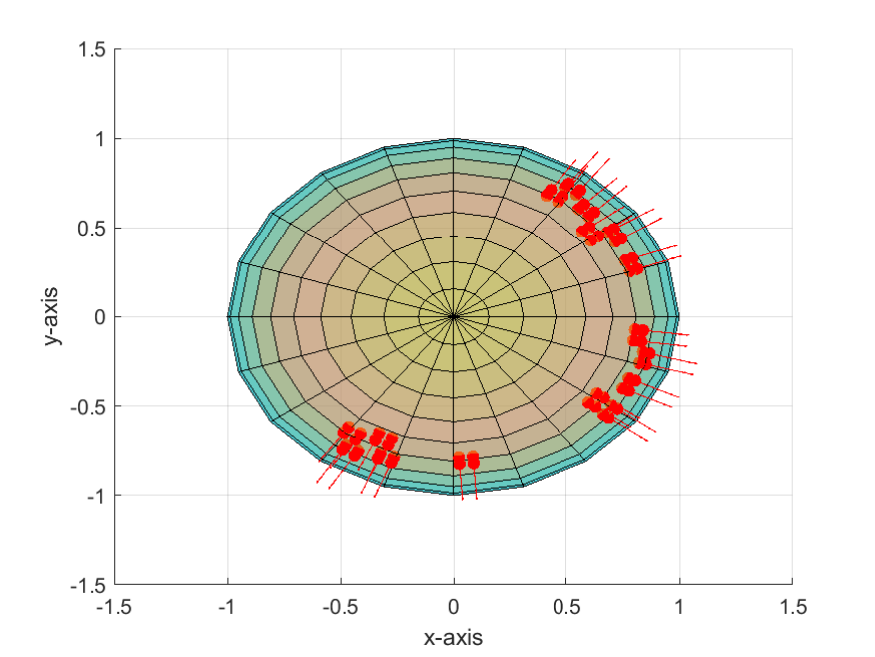}}
		\subfigure[HT-6DMA.]{ \label{fig:Airway_sensing_opt_SMA_3D}
			\includegraphics[width=1.7in]{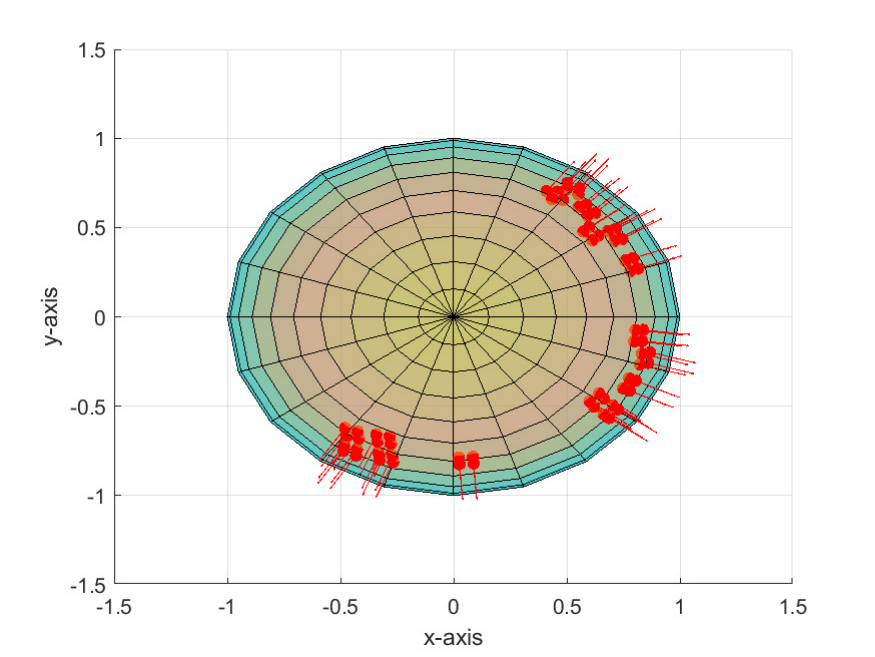}}	
		\subfigure[Beampattern of FPA.]{ \label{fig:Airway_sensing_FPA_beampattern}
			\includegraphics[width=1.7in]{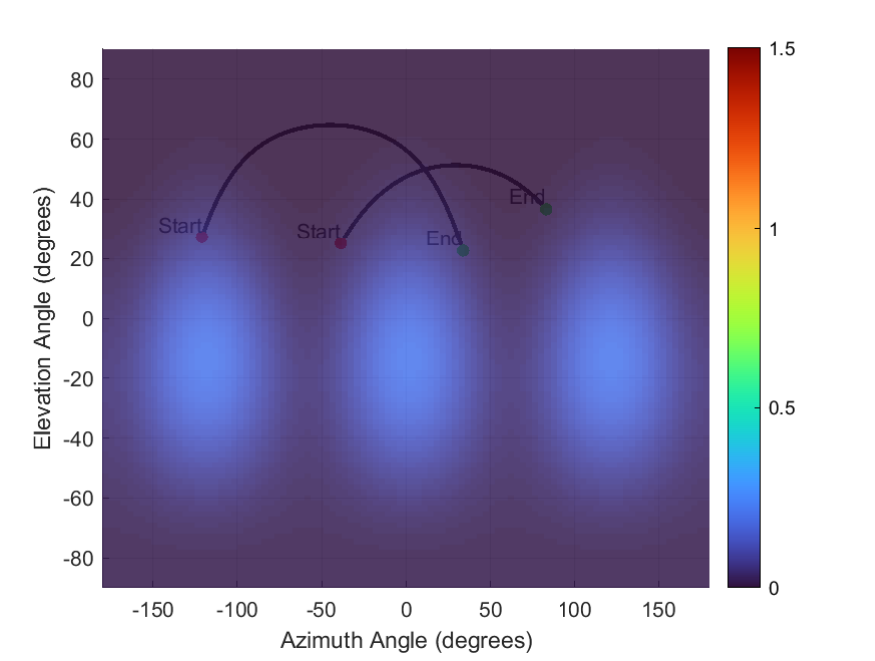}}
		\subfigure[Beampattern of HT-6DMA with RA.]{  \label{fig:Airway_sensing_opt_Rot_beampattern} % Add FPA is left later
			\includegraphics[width=1.7in]{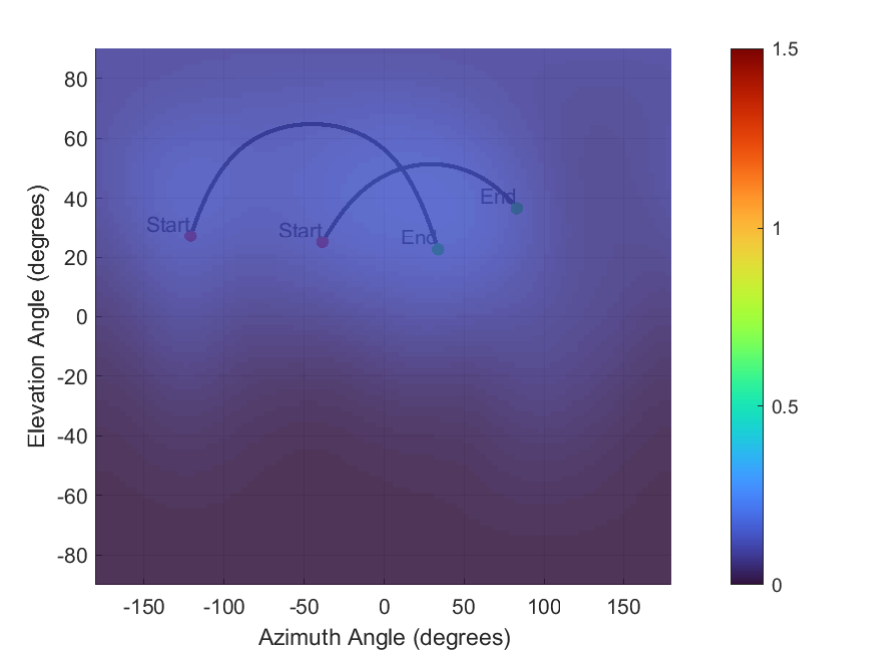}}
		\subfigure[Beampattern of HT-6DMA.]{ \label{fig:Airway_sensing_opt_SMA3D_beampattern}
			\includegraphics[width=1.7in]{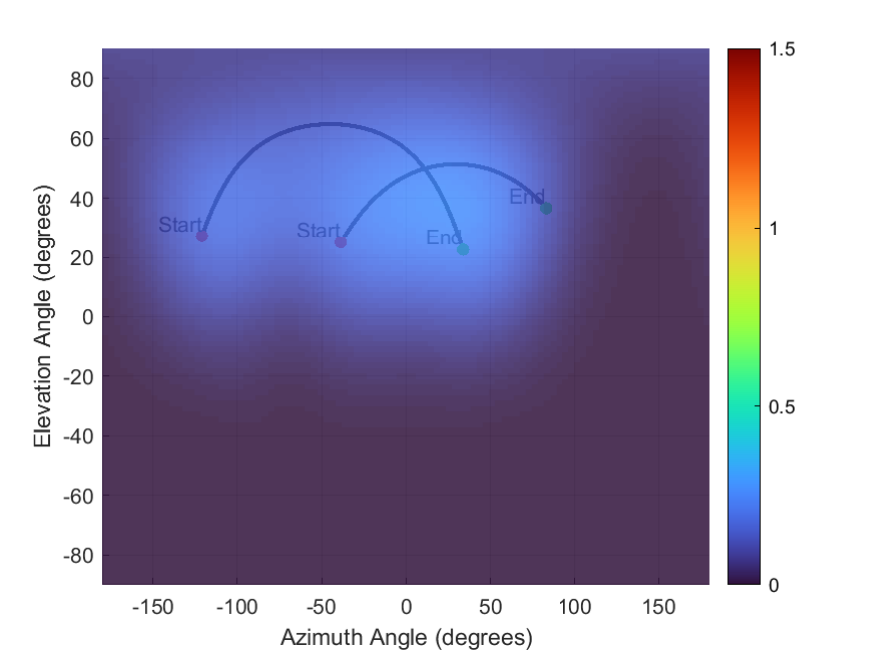}}
		\subfigure[Beampattern of HT-6DMA with optimized $\bm{R}_d$.]{ \label{fig:Airway_sensing_optRd_SMA3D_beampattern}
			\includegraphics[width=2.053in]{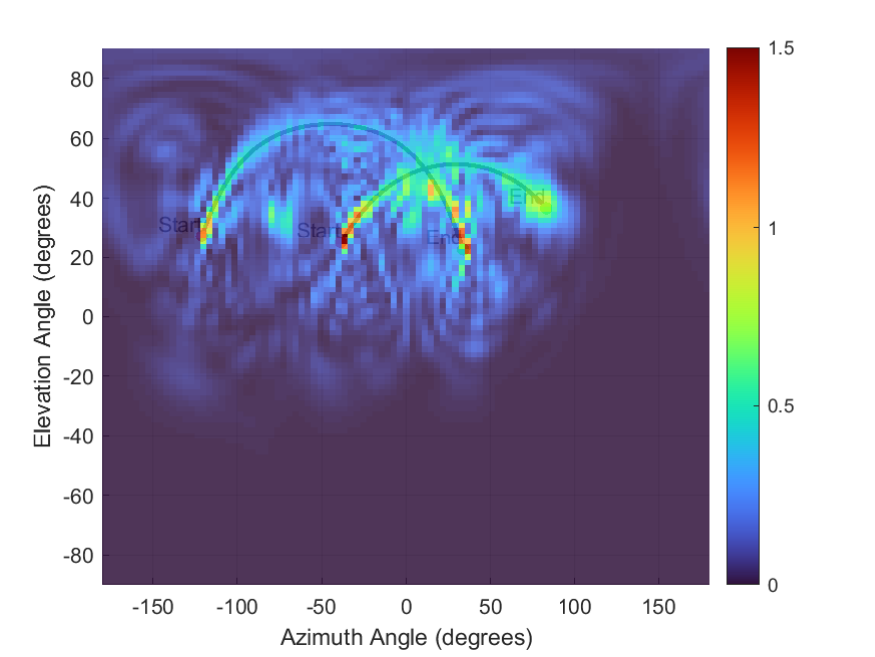}}
		\subfigure[Received power along the first airway segment.]{ \label{fig:Beampattern_1way}
			\includegraphics[width=2.053in]{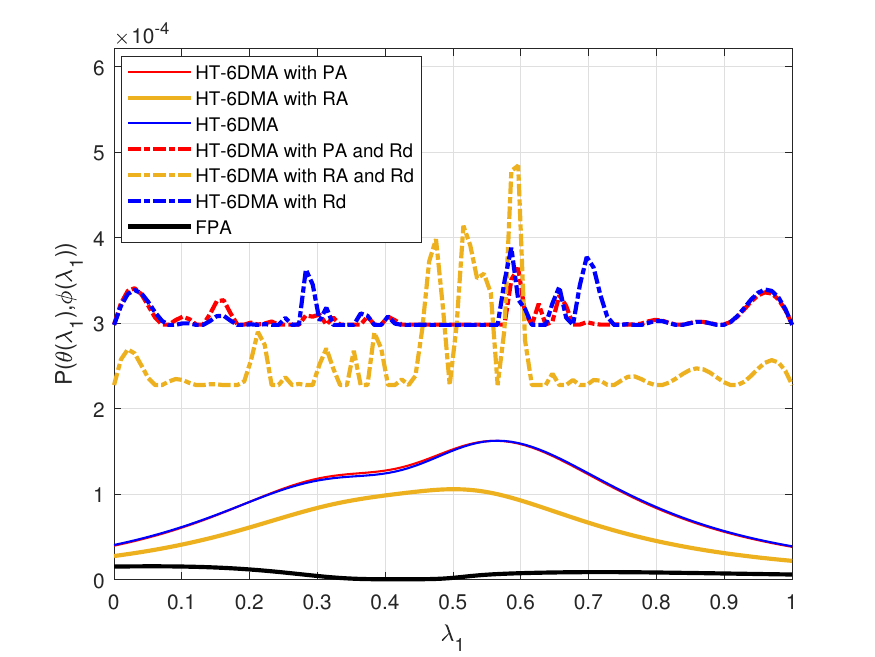}} % 2.4in
		\subfigure[Received power along the second airway segment.]{ \label{fig:Beampattern_2way}
			\includegraphics[width=2.053in]{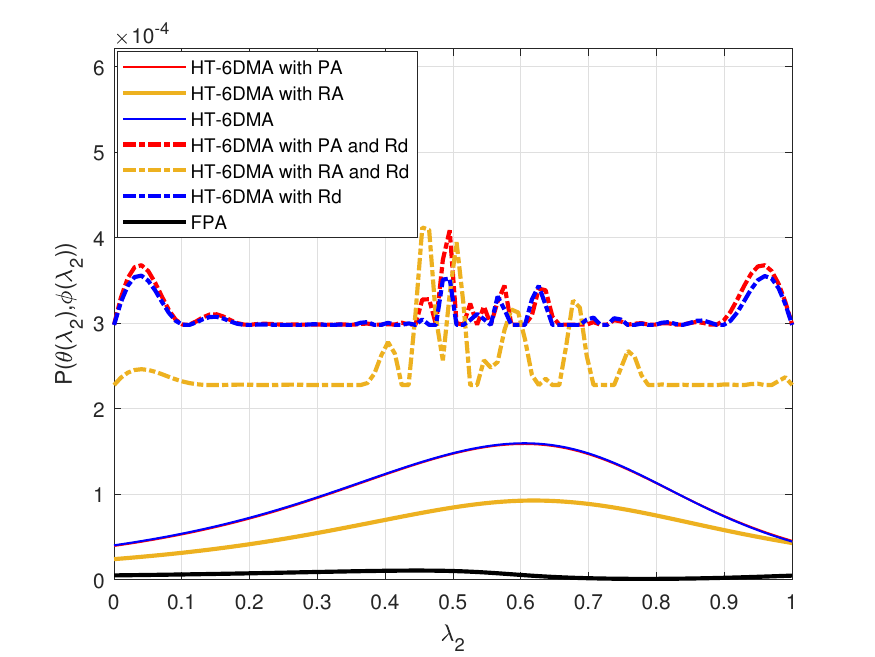}} % 2.4in
		\caption{Simulation and optimization results in airway sensing scenario.}
		\label{fig:Sensing_scenario}
	\end{figure*}
	Fig. \ref{fig:Cap_vs_xi} shows the achievable sum rate versus the user homogeneous ratio $\eta$ with $B=16$ and $N =   4$. It is observed that when $\eta$ increases from 0 to 1, the sum rate of the FPA scheme increases while those of all the HT-6DMA and existing 6DMA schemes decrease. This is due to the fact that when the users are more homogeneously/uniformly distributed (i.e., with larger $\eta$), the benefit of 6DMA in flexibly allocating the antenna resources for hotspot areas becomes less significant. 
	%	For FPA, even if the 3 sectors are fixed with their position and orientation, more users can be properly served when $\eta$ becomes large.
	It is also observed that when the user is more homogeneously/uniformly distributed, the performance gaps among different 6DMA schemes become smaller.
	%	  as when $\eta$ is large, proper design of the arrays becomes less important due to the lack of user clusters. 
	
	Fig. \ref{fig:Gap_d_min} shows the sum rate performance comparison between the HT-6DMA (with both PA and RA) and the HT-6DMA with PA under different values of the inter-array minimum distance $d_{\text{min}}$, where their performance differences characterize the gain of array local rotation with HT-6DMA. The setup in Fig. \ref{fig:Gap_d_min} is the same as that in Fig. \ref{fig:uplink_comm_scenario} for each $d_{\text{min}}$.
	First, it is observed from Fig. \ref{fig:Gap_d_min} that with larger $d_{\text{min}}$, the sum rates of both HT-6DMA schemes decrease due to more strigent inter-array spacing requirement. 
	It is also observed that when $d_{\text{min}}$ increases, the performance gain by rotation adjustment also increases.  
	This is because when $d_{\text{min}}$ becomes large, it is more difficult to position all arrays on the spherical surface such that their default orientations can point accurately towards the corresponding hotspot area.
	As such, the additional local orientation/rotation adjustment of each array towards the hotspot areas can bring higher performance gain.

	\subsection{Airway Sensing Scenario}\label{sec:sensing_numerial}
	
	Next, we present the simulation results
	for the airway sensing scenario in Fig. \ref{fig:Sensing_scenario}, where $\beta$ is set to be $50$. Specifically, we consider two airway segments intersecting with each other as shown in Fig. \ref{fig:Airway_3D}, where the first one starts at $\bm{r}_s^1 = \left[-30, -50, 30\right]^T$ m and ends at $\bm{r}_e^1 = \left[60, 40, 30\right]^T$ m and the second one starts at $\bm{r}_s^2 = \left[50, -40, 30\right]^T$ m and ends at $\bm{r}_e^2 = \left[5, 40, 30\right]^T$ m. 
	
	Fig. \ref{fig:Convergence_curve_sensing} shows the convergence behaviors of different HT-6DMA algorithms in terms of the surrogate function $f_a$, with the same initial variable values as those shown in Fig. \ref{fig:6DMA_init_2}. It is observed that most of the improvement in terms of $f_a$ is achieved by the array position adjustment on the spherical surface instead of the local rotation adjustment, which is consistent with the uplink communication scenario.
	Fig. \ref{fig:Airway_sensing_opt_SMA_1D} - Fig. \ref{fig:Airway_sensing_opt_SMA_3D} show the optimized array positions and rotations viewed from above for different HT-6DMA schemes\footnote{Note that the existing 6DMA algorithm proposed in \cite{shao20246d} is not considered for performance comparison in the airway sensing scenario because it is found that adjusting the array position with its global orientation fixed cannot improve the objective function of problem (P2).}. % Yes, it can be shown, but you do not show it, in the simulation, you just find it.. Add in the revision.
	It is observed in Fig. \ref{fig:Airway_sensing_opt_SMA} and Fig. \ref{fig:Airway_sensing_opt_SMA_3D} that the antenna arrays are all oriented towards the two airway segments, showing the effectiveness of the proposed algorithms based on the HT-6DMA model. 
	Moreover, compared with Fig. \ref{fig:Airway_sensing_opt_SMA}, Fig. \ref{fig:Airway_sensing_opt_SMA_3D} shows that the arrays are further rotated to improve the minimum received power along the airways. 
	In contrast, Fig. \ref{fig:Airway_sensing_opt_SMA_1D} shows that although some arrays are optimized to face towards the airways, other arrays fail to achieve that due to their fixed positions and the practical constraints imposed on their orientation adjustments.
	
	To compare the sensing performance achieved by different schemes, we show the received power distribution (beampattern) in the 2D angular plot for
	the FPA, HT-6DMA with RA, and HT-6DMA (with both PA and RA)\footnote{The beampattern of the HT-6DMA with PA is similar to that of the HT-6DMA (with both PA and RA) and is thus omitted.} in Fig. \ref{fig:Airway_sensing_FPA_beampattern}, Fig. \ref{fig:Airway_sensing_opt_Rot_beampattern}, and Fig. \ref{fig:Airway_sensing_opt_SMA3D_beampattern}, respectively, with the transmit covariance matrix $\bm{R}_d$ not optimized yet (i.e., $\bm{R}_d = \frac{P_0}{NB} \bm{I}_{NB}$).
	The HT-6DMA (with both PA and RA) achieves the highest minimum sensing signal power over the two airway segments, demonstrating the effectiveness of our proposed algorithm for array position/rotation optimization.
	Moreover, if $\bm{R}_d$ is further optimized in the proposed HT-6DMA algorithm, it is observed from Fig. \ref{fig:Airway_sensing_optRd_SMA3D_beampattern} that the signal power is more focused along the airway segments, leading to enhanced sensing performance compared with Fig. \ref{fig:Airway_sensing_opt_SMA3D_beampattern}.
	
	Under the same setup, we further show the received power by different designs along the first and second airway segments in Fig. \ref{fig:Beampattern_1way} and Fig. \ref{fig:Beampattern_2way}, respectively, as well as their corresponding minimum received power achieved in Table \ref{tab:achieved_chi}.
	These results are consistent with the beampatterns for different schemes shown in Figs. \ref	{fig:Airway_sensing_FPA_beampattern} - \ref{fig:Airway_sensing_optRd_SMA3D_beampattern}.
	It is observed that, similar to the uplink communication scenario, the performance
	\begin{table}[htb]
		\centering
		\caption{Minimum received power comparison}
		\label{tab:achieved_chi}
		\begin{tabular}{lc}
			\toprule
			\textbf{Configuration} & \textbf{Minimum received power} \\
			& \textbf{(mW)} \\
			\midrule
			Initial setup              & 0.00110 \\
			HT-6DMA with PA          & 0.00384 \\
			HT-6DMA with RA          & 0.00220 \\
			HT-6DMA                 & 0.00391 \\
			HT-6DMA with PA and (optimized) $\bm{R}_d$   & 0.02985 \\
			HT-6DMA with RA and (optimized) $\bm{R}_d$   & 0.02277 \\
			HT-6DMA with (optimized) $\bm{R}_d$ & 0.02981 \\
			FPA                     & 0.00005 \\
			\bottomrule
		\end{tabular}
	\end{table}
	gain by HT-6DMA in the sensing scenario also mostly comes from the array position optimization rather than the array local rotation optimization as the performance gaps between the HT-6DMA with PA schemes and the HT-6DMA (with both PA and RA) schemes are very small, as shown in Table \ref{tab:achieved_chi}.
	
	\begin{figure}[t]
		\centering
		\includegraphics[width=2.1in]{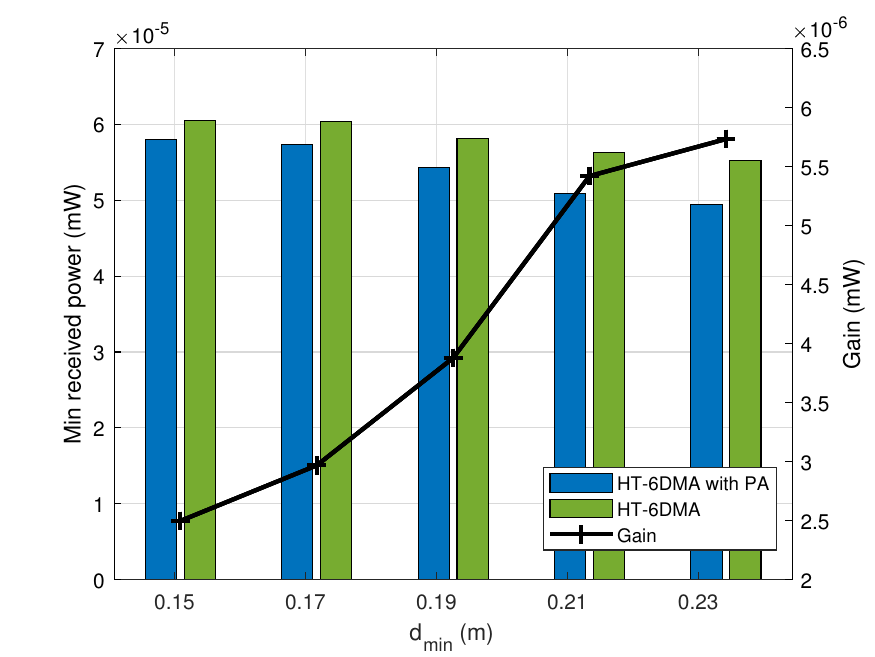}
		\centering
		\caption{The minimum received power performance comparison and the gain of array local rotation with HT-6DMA versus $d_{\text{min}}$.}
		\label{fig:Gap_d_min_s}
	\end{figure}
	
	Finally, Fig. \ref{fig:Gap_d_min_s} shows the minimum recieved signal power performance comparison between the HT-6DMA (with both PA and RA) and the HT-6DMA with PA with various $d_{\text{min}}$, where
	their performance differences characterize the sensing gain of array local rotation with HT-6DMA. For simplicity, we only consider the second airway here. Similar to the uplink communication scenario, it is observed from Fig. \ref{fig:Gap_d_min_s} that with larger $d_{\text{min}}$, the sensing performance of both design schemes decrease while the performance gain by local rotation adjustment increases.

	%		\begin{figure}[t]
		%				\centering
		%				\setlength{\abovecaptionskip}{+4mm}
		%				\setlength{\belowcaptionskip}{+1mm}
		%				\subfigure[The additional gain versus $d_{\text{min}}$.]{ \label{fig:Gap_d_min_s}
			%					\includegraphics[width=1.6725in]{./figures/Gap_d_min_s.eps}}
		%				\subfigure[The additional gain versus the shape of the UPA.]{ \label{fig:Gap_shape_s}
			%					\includegraphics[width=1.6725in]{./figures/Gap_shape_s_enhanced.eps}}
		%				\caption{The additional sensing performance gain of the HT-6DMA over the HT-6DMA with PA versus $d_{\text{min}}$ and shape of UPA, respectively.}
		%				\label{fig:Sensing_gap_comp}
		%		\end{figure}
	
	\section{Conclusion}\label{sec:conclusion}
	
	This paper proposed a new HT-6DMA BS architecture for future wireless networks to improve the performance of both wireless communication and sensing over conventional FPAs. The HT-6DMA BS consists of multiple arrays equipped with directional antennas, where each array can flexibly move on a spherical surface, with their 3D positions and 3D rotations/orientations efficiently characterized in the global SCS
	and their individual local SCSs, respectively. The proposed HT-6DMA is hierarchically tunable in the sense that each array’s position and rotation can be individually adjusted in a sequential manner with the other being fixed, thus greatly reducing their optimization complexity and improving the solution quality. 
	To validate its effectiveness, we considered both the multi-user uplink communication scenario and the airway sensing scenario. In particular, for communication scenario, we maximized the average sum rate of all users in the long term via optimizing the positions and orientations of all 6DMA arrays at the BS subject to practical constraints. 
	While for airway sensing, we maximized the minimum received power along the airway segments via optimizing both the array positions/rotations as well as the transmit signal covariance subject to practical constraints. Leveraging the hierarchical tunability of position/rotation of 6DMA arrays in our proposed model, we proposed efficient algorithms to solve the considered problems for both communication and sensing scenarios.
	Numerical results showed that the proposed designs can significantly enhance the performance in both communication and sensing compared to conventional FPAs. It was also shown that the proposed algorithms based on the new HT-6DMA model converge to better solutions within fewer iterations than the existing 6DMA model based algorithm in communication scenario. Furthermore, under both considered scenarios, it was revealed that the performance gains of HT-6DMA mostly come from the arrays' global position adjustments on the spherical surface instead of their local rotation adjustments. 
	Nevertheless, it was also found that the additional gain achieved by arrays' local rotation adjustments increases when the required minimum array spacing increases. These findings provide useful guidance for implementing 6DMA systems with a balanced performance-complexity trade-off in practice.

		\bibliographystyle{IEEEtran}
		
		%	\bibliographystyle{IEEEbib}
		%% Put references in BibTeX format in refs.bib.
		%% since for some reference, we need the abbreviation of the conference/journal/Transaction, first check it in BibGuru, then just change the full name of the conference/journal/Transaction into its corresponding abbreviation.
		\bibliography{refsv3}
		%%%%%%%%%%%%%%%%%%%%

	\end{document}